\documentclass[fleqn,usenatbib]{mnras}

\usepackage{newtxtext,newtxmath}

\usepackage[T1]{fontenc}
\usepackage{url}
\usepackage{mathtools}
\usepackage[normalem]{ulem}
\usepackage{xcolor}
\DeclareRobustCommand{\VAN}[3]{#2}
\let\VANthebibliography\thebibliography
\def\thebibliography{\DeclareRobustCommand{\VAN}[3]{##3}\VANthebibliography}

\usepackage{graphicx}	
\usepackage{amsmath}	

\title[Reducing Weak-lensing Errors on Standard Sirens]{Reducing the Impact of Weak-lensing Errors on Gravitational-wave Standard Sirens}

\author[Z. Wu et al.]{
Zhao-Feng Wu,$^{1}$\thanks{E-mail:
Foisonwzf@link.cuhk.edu.hk}
Lok W. L. Chan,$^{1}$\thanks{E-mail:
lokwlc@link.cuhk.edu.hk}
Martin Hendry,$^{2}$\thanks{E-mail:
martin.hendry@glasgow.ac.uk}
Otto A. Hannuksela$^{1}$\thanks{E-mail:
hannuksela@phy.cuhk.edu.hk}
\\
$^{1}$Department of Physics, The Chinese University of Hong Kong, Shatin, New Territories, Hong Kong\\
$^{2}$SUPA, School of Physics and Astronomy, University of Glasgow, UK}

\date{Accepted XXX. Received YYY; in original form ZZZ}

\pubyear{2022}

\begin{document}
\label{firstpage}
\pagerange{\pageref{firstpage}--\pageref{lastpage}}
\maketitle

\begin{abstract}
The mergers of supermassive black hole binaries (SMBHBs) can serve as standard sirens: the gravitational wave (GW) analog of standard candles. The upcoming space-borne GW detectors will be able to discover such systems and estimate their luminosity distances precisely. Unfortunately, weak gravitational lensing can induce significant errors in the measured distance of these standard sirens at high redshift, severely limiting their usefulness as precise distance probes. The uncertainty due to weak lensing can be reduced if the lensing magnification of the siren can be estimated independently, a procedure called 'delensing'. With the help of up-to-date numerical simulations, here we investigate how much the weak-lensing errors can be reduced using convergence maps reconstructed from shear measurements. We also evaluate the impact of delensing on cosmological parameter estimation with bright standard sirens. We find that the weak-lensing errors for sirens at $z_s = 2.9$ can be reduced by about a factor of two on average, but to achieve this would require expensive ultra-deep field observations for every siren. Such an approach is likely to be practical in only limited cases, and the reduction in the weak-lensing error is therefore likely to be insufficient to significantly improve the cosmological parameter estimation. We conclude that performing delensing corrections is unlikely to be worthwhile, in contrast to the more positive expectations presented in previous studies. For delensing to become more practicable and useful in the future will require significant improvements in the resolution/depth of weak-lensing surveys and/or the methods to reconstruct convergence maps from these surveys.
\end{abstract}

\begin{keywords}
	gravitational lensing: weak -- gravitational waves -- distance scale -- cosmological parameters

\end{keywords}



\section{Introduction}
\setcounter{footnote}{-1}
\begin{figure*}
    \centering
    \includegraphics[width=\linewidth]{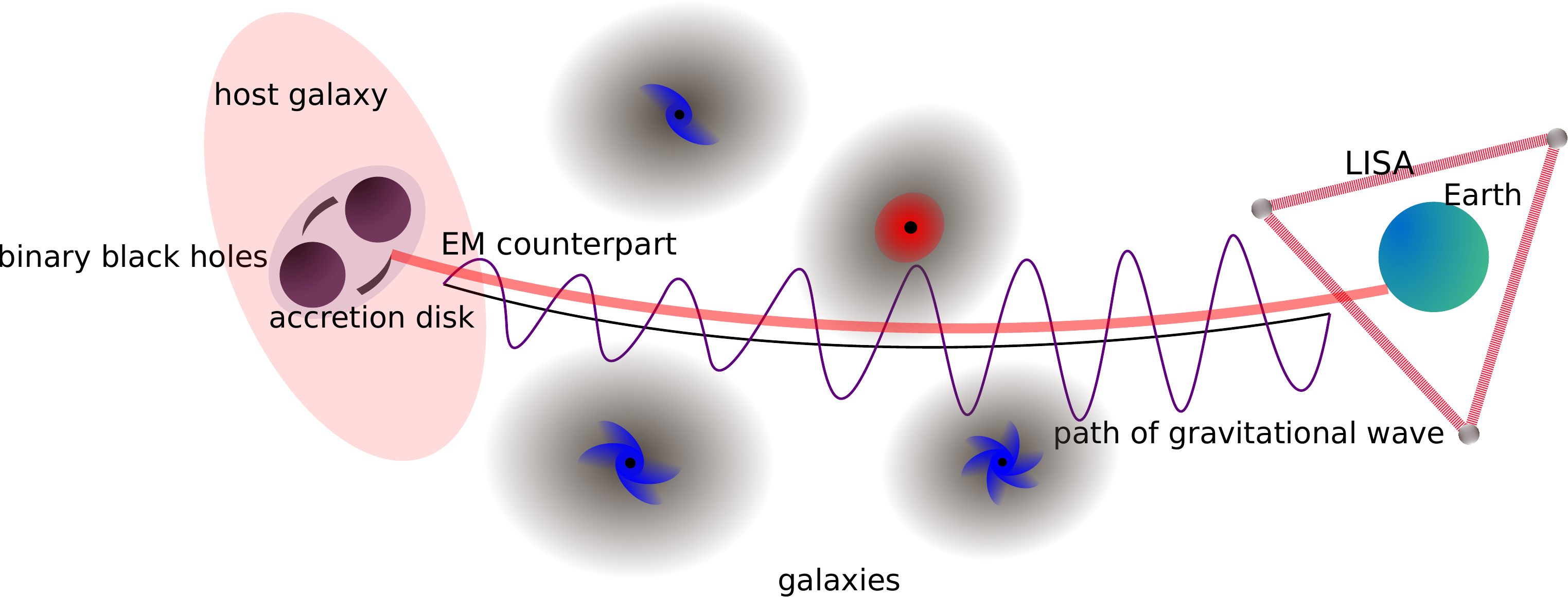}
    \caption{An illustration of the effect of weak gravitational lensing on the GWs emitted by binary black hole systems (SMBHBs in our case). We show here both the host galaxy of the binary and the circumbinary accretion disk\protect\footnotemark that may be responsible for the EM counterpart. The dark matter haloes around the intervening galaxies will (de)magnify the GW signals, adding uncertainties to the amplitude of the original waveforms. The GW signals will be measured by space-borne GW detectors (\textit{LISA} in the illustration as an example) and the EM counterpart will be observed by electromagnetic telescopes.}
    \label{fig:illustration}
\end{figure*}
A promising method to obtain accurate cosmological distance measurements in the future is to observe gravitational waves emitted by merging supermassive black hole binaries (SMBHBs). \citet{schutzDeterminingHubbleConstant1986} showed that measuring the GW signals of the inspiraling binary with a network of interferometers could estimate its luminosity distance, independently of the cosmic distance ladder. That is why these binaries are coined as 'standard sirens' -- the gravitational wave analog of standard candles. 

However, the redshift cannot be measured from the gravitational waves alone, so identifying an electromagnetic (EM) counterpart is also crucial for cosmological purposes \citep{dalalShortGRBBinary2006}. Such an electromagnetic counterpart would exist if the merging source was a binary neutron star (BNS), in which case a potentially detectable gamma-ray burst would be associated with the merger \citep{2007PhR...442..166N}, or may exist if the merging source was an SMBHB, in which case the accretion disk surrounding the black holes may give off an electromagnetic signal \citep{2005ApJ...622L..93M,2012MNRAS.420..705T,yuanPostmergerJetsSupermassive2021}. 
If the EM counterpart of the coalescence can be observed, and thus the redshift of the host galaxy determined, the cosmological parameters can then be estimated by analyzing the relationship between the luminosity distance and the redshift of these standard sirens.

In this work, only GWs generated by the coalescence of supermassive (or massive) black hole binaries are considered. These systems are expected to be luminous in GWs such that these systems may be detectable even up to $z=5$. However, the frequency of the expected GW signals from SMBHBs lies below the sensitive band of ground-based GW detectors and is thus better suited for space-based detection. 

Scientists have proposed several space-borne GW detector projects to extend the GW spectrum to the millihertz band, such as \textit{Laser Interferometer Space Antenna} (\textit{LISA}) \citep{amaro-seoaneLaserInterferometerSpace2017} and \textit{TianQin} (\textit{TQ}) \citep{luoTianQinSpaceborneGravitational2016}. GW detectors in space benefit from their long baselines and the absence of seismic noise. A baseline of 2.5 million kilometers \citep{2019BAAS...51g..77T} between spacecraft allows the detection of GWs in the millihertz band. 

The expected number and redshift distribution of SMBHBs that \textit{\textit{LISA}} and \textit{TQ} will observe are very uncertain and strongly dependent on details of our model for the mergers of galaxy nuclei \citep{Arun_2009}. However, observing even a handful of SMBHBs could constrain the standard cosmological model with impressive accuracy, as was first demonstrated by \citet{holzUsingGravitationalWave2005}. The expected measurement errors of the luminosity distance from the SMBHB merger can be dramatically small compared with the scatter present for other cosmological probes. At the same time, many SMBHBs are anticipated to be detected at high redshift. These data will, therefore, potentially be of great value for the measurement of cosmological parameters such as $H_0$ and the dimensionless densities of dark matter and dark energy.

However, as was also pointed out in \citet{holzUsingGravitationalWave2005}, there is a huge caveat: GW signals from the SMBHBs would be affected by weak gravitational lensing, which is caused by fluctuations in the density of matter along the line of sight to the siren. When the sky position of the targeted siren coincides with a foreground galaxy or cluster, then strong lensing takes place, which can provide much more information than a standard siren alone \citep{pangLensedNotLensed2020a,Hannuksela2020LocalizingLensing}. While such strong lensing events will be very useful, their probability of occurring is extremely small. Besides, the uncertainties in the lens model also limit their usefulness. Consequently, only weak gravitational lensing will be considered in this work. Weak gravitational lensing is caused by extended dark matter halos that lie along the line of sight to the siren; the effect is weaker yet more prevalent for high-redshift standard sirens like SMBHBs (See Fig.~\ref{fig:illustration} for illustration). The error induced on the luminosity distance by the weak-lensing effect is expected to be a few percent for sources at high redshift, which is comparable to the scatter of the SMBHB siren luminosity distance estimate itself \citep{Holz_2005,Kocsis_2006}. Therefore, the weak-lensing effect substantially limits the power of the SMBHB sirens as precise cosmological distance probes. 

\footnotetext{We neglect possible circum-single disks around each black hole and a probable binary cavity in the illustration, which may also help identify the EM counterparts.}

To improve the performance of SMBHB sirens, corrections from EM observations may be applied to reduce the impact of weak-lensing errors. \citet{shapiroDelensingGravitationalWave2010b} proposed a method based on estimating the weak-lensing signal in the direction of each siren using maps of the shear and flexion reconstructed from surveys of foreground galaxies. In this way we can infer the weak-lensing error affecting each siren from the reconstructed maps and hence correct for that error on the siren's luminosity distance estimate -- a procedure called ‘delensing’. 

In previous studies of delensing \citep{shapiroDelensingGravitationalWave2010b,10.1111/j.1365-2966.2010.17963.x}, the authors assumed a model for the matter fluctuations across cosmic time either from analytical formulae or numerical simulations. In \citet{shapiroDelensingGravitationalWave2010b}, the analytical formulae used had been extended beyond their accepted range of accuracy to reach the required higher resolution. On the other hand, the numerical simulations used in \citet{10.1111/j.1365-2966.2010.17963.x} had an adequate resolution, but the values of the cosmological parameters adopted by the simulations deviate from the more recently accepted values.\footnote{The cosmological parameters used in the backbone N-body simulations in \citet{10.1111/j.1365-2966.2010.17963.x} are:  $\Omega_m = 0.25$,  $\Omega_\Lambda = 0.75$,  $h = 0.73$,  $n = 1$,  $\sigma_8 = 0.9$.} As pointed out in \citet{10.1111/j.1365-2966.2010.17963.x}, the optimal smoothing strategy and outcomes of delensing should be cosmologically dependent, so a revisit is necessary to investigate the reliability of the predictions made in that work.

Now with the availability of up-to-date numerical simulations, a more complete and consistent treatment can be applied to evaluate the potential of delensing by weak-lensing reconstruction. Compared to the previous studies, we focus on making the evaluation more realistic, and more in line with the expected performance of future galaxy surveys. Due to the uncertainties in the advances of technology, here we adopt optimistic settings for the measurement uncertainties to explore the optimal ability of delensing in the future.
Moreover, in this paper we also use up-to-date SMBHB siren mock catalogs, allowing an explicit investigation of the impact of weak-lensing error reduction on cosmological parameter estimation. However, we do not perform a joint standard siren + weak gravitational lensing analysis, which has been discussed in other studies \citep{congedoJointCosmologicalInference2019a,2022arXiv220805959M,2022arXiv221006398B}. We also do not exploit the non-Gaussianity of the lensing distributions.

The goal of this paper is, therefore, to evaluate the potential of weak-lensing reconstruction for reducing weak-lensing errors on SMBHB sirens, and to investigate the impact on cosmological parameter estimation. We attempt to strike a balance between optimistic and realistic assumptions on future surveys so that the predictions could be more useful and reliable in the future. In Sec.~\ref{general} and ~\ref{sec:wek_recon}, we introduce the necessary background information for standard siren cosmology and weak-lensing reconstruction. A Bayesian analysis to investigate the impact of delensing on cosmological parameter estimation is then described in Sec.~\ref{baye}, and the construction of the best estimator of weak-lensing error is presented in Sec.~\ref{estimator}. In Sec.~\ref{sim}, we introduce the numerical simulations adopted in our analysis. In Sec.~\ref{result}, we compute the reduction in the weak-lensing error under different observational schemes. We discuss the outcomes of the previous delensing studies and compare them to ours in Sec.~\ref{Discussion}. Finally, we conclude in Sec.~\ref{Conclusion} along with a discussion about topics for further investigation. 

\section{Methodology}\label{method}
\subsection{Standard siren cosmology}\label{general}
\subsubsection{Standard sirens}
For the inspiral of a compact binary system with component masses $m_1$ and $m_2$, the frequency domain GW waveform can be expressed as \citep{sathyaprakashPhysicsAstrophysicsCosmology2009},
\begin{equation}\label{waveform}
\widetilde{h}(f)= \frac{1}{D_L} \sqrt{\frac{5}{24}} \frac{(G\bar{M_z})^{5/6}}{\pi^{2/3}c^{3/2}} f^{-7/6} \exp(-i\Phi(f;\bar{M_z}, \eta))
\end{equation}
where $G$ is the gravitational constant, $\bar{M} = \eta^{0.6}M$ is the chirp mass, $\eta = m_1 m_2/M^2$ is the symmetric mass ratio, $M= m_1+m_2$ is the total mass, and $\Phi$ is phase of the waveform. The source luminosity distance $D_L$ directly appears in the waveform, which provides luminosity distance information free from the ‘cosmic distance ladder’. Therefore, these compact binary systems are coined as ‘standard sirens’ and their measured luminosity distances may have small systematic and statistical errors compared to other approaches. One type of standard siren is the SMBHB system that generates GW signals detectable by space-borne GW detectors. We refer to SMBHB sirens simply as (standard) sirens hereafter.

However, according to Eq.~\eqref{waveform}, only the redshifted chirp mass $\bar{M_z} \equiv \bar{M}(1+z)$ is accessible in the waveform, and the redshift $z$ is therefore degenerate with the chirp mass $\bar{M}$.  Consequently, measurements of redshift must rely on extra information. In this work, the EM counterparts of the GW signals are assumed to be accessible so that the degeneracy between redshift and mass is broken -- i.e. a scenario where the sirens are ‘bright’. 

Possible EM counterparts from SMBHBs include the broadband nonthermal EM emission from electrons accelerated at the external forward shock expected in post-merger relativistic jets from the coalescence \citep{yuanPostmergerJetsSupermassive2021}. If the binary’s tidal torques are able to open a central cavity in the accreting gaseous disc, the SMBHB might have an unusually low soft X-ray luminosity and weak ultraviolet and broad optical emission lines, as compared to an AGN powered by a single supermassive black hole with the same total mass \citep{2012MNRAS.420..705T}. This could also help identify the EM counterparts of the sirens.

\subsubsection{Estimating cosmological parameters}
In this paper, we focus on the estimation of dimensionless density parameters of dark matter and dark energy respectively, labelled $\Omega_M$ and $\Omega_\Lambda$, without assuming a flat universe\footnote{However, the simulations used in this paper all assumed {\em true\/} values of the cosmological parameters consistent with a flat universe, i.e. $\Omega_K = 0$, when generating the data.} under the $\Lambda CDM$ model. This is because the SMBHBs are detectable even for redshift up to $z = 3$, where the difference in the expansion history becomes more distinguishable for different values of $\Omega_\Lambda$ and $\Omega_m$ compared to the low-redshift case.

Once the luminosity distance $D_L$ and redshift $z$ of each siren have been measured, we can use the theoretical relation between $D_L$ and $z$, which depends on the expansion history of the universe, to estimate the cosmological parameters. The theoretical relationship is given by,
\begin{equation}\label{lum_z}
D_L = \frac{c(1+z)}{H_0}\begin{dcases}
\frac{1}{\sqrt{\Omega_K}} \sinh(\sqrt{\Omega_K}\int^z_0 \frac{H_0}{H(z')} \, dz') & \text{for $\Omega_K$ > 0}\\
\int^z_0 \frac{H_0}{H(z')} \, dz' & \text{for $\Omega_K$ = 0}\\
\frac{1}{\sqrt{|\Omega_K|}} \sin(\sqrt{|\Omega_K|}\int^z_0 \frac{H_0}{H(z')} \, dz') & \text{for $\Omega_K$ < 0}
\end{dcases}
\end{equation}
where $H_0 \equiv H(z = 0)$ describes the current expansion rate of the Universe and c is the speed of light in vacuum. The Hubble constant for different redshifts $H(z)$ is governed by the normalized Friedmann equation,
\begin{equation}
H(z) = H_0\sqrt{\Omega_M(1+z)^3 +\Omega_K(1+z)^2 + \Omega_\Lambda},
\end{equation}
where $\Omega_M$, $\Omega_K$, and $\Omega_\Lambda$ are fractional densities of the total matter, curvature, and dark energy with respect to the critical density $\rho_c = 3{H^2_0}/8\pi G$ respectively.

The observed luminosity distance of each siren ${D_L}^{\rm obs}$ is subject to various sources of noise that include measurement uncertainties, the impact of host galaxy peculiar velocities and the effect of weak gravitational lensing. Measurement noise could be reduced by advances in the relevant GW detector technologies, while the SMBHBs are assumed to be distant enough so that the effect of peculiar velocity is negligible. In this work, we concentrate on the weak-lensing effect.

\subsubsection{Weak gravitational lensing \& delensing}\label{sec:wl&delensing}
Unlike the impact of peculiar velocities, which will reduce with distance, the uncertainties due to weak gravitational lensing become more dominant for high-redshift sirens \citep{hirataReducingWeakLensing2010a}. GWs generated from those sources have a larger probability to be significantly lensed by the intervening dark matter halos as they propagate through the universe. Therefore, reducing the errors from weak gravitational lensing for the high-redshift sirens should be important. A brief introduction about the weak-lensing effect and the idea of ‘delensing’ is as follows. A more detailed and technical introduction is given in Section~\ref{sec:wek_recon}.

In the weak-lensing and geometrical optics limit, the observed luminosity distance to a siren is related to its true luminosity distance by 
\begin{equation} \label{eq:1}
{D_{L}}^{\rm obs}=D_{L}/\sqrt{\mu}\:\approx\:D_{L}(1-\delta\mu/2) \,,
\end{equation}
where $\mu$ is the lensing magnification and $\delta \mu=\mu-1$ is the deviation from the unlensed case \citep{holzUsingGravitationalWave2005,Takahashi_2006}. 
Note that $\delta \mu=0$ if the siren is not lensed. If the lensing magnification experienced by each siren can be estimated from other measurements, then we can compensate for the weak-lensing errors and obtain corrected observed luminosity distances $D_{L}^{\rm obs}$. This is the fundamental motivation of this work and the origin of the name 'delensing'.

\subsection{Bayesian analysis on impact of delensing}\label{baye}

In this section, we introduce a Bayesian analysis to quantify the impact of delensing on cosmological parameter estimation by standard sirens. The posterior on the cosmological parameters is given by\footnote{See Appendix~\ref{appd} for a detailed derivation.}
\begin{equation}
p(\Vec{\Omega}|\,\mathbf{D}_{\rm GW}\,, \mathbf{z}_{\rm s}\,,\mathbf{d}_{\rm lens}) \propto p(\Vec{\Omega}|\, \mathbf{z}_{\rm s}\,,\mathbf{d}_{\rm lens}) p(\mathbf{D}_{\rm GW}\,|\,\Vec{\Omega}, \mathbf{z}_{\rm s}\,,\mathbf{d}_{\rm lens}) \,,
\end{equation}
where $\Vec{\Omega}$ is e.g. a two-dimensional vector consisting of $\Omega_M$ and $\Omega_\Lambda$, $\mathbf{D}_{\rm GW}$ is the vector of GW data, $\mathbf{z}_{\rm s}$ is the vector of siren's redshifts determined by EM counterparts, $\mathbf{d}_{\rm lens}$ is the lensing data used for delensing. In this paper, the lensing data $\mathbf{d}_{\rm lens}$ include the sky positions, observed ellipticities, and redshifts of galaxies.

Note that here we are assuming that $H_0$ is already known and fixed in value; we could also consider the joint inference of $H_0$ and the other cosmological parameters, but we expect that the $H_0$ tension will be resolved and $H_0$ can be approximated as fixed by the time \textit{LISA} operates\footnote{As pointed out by \citet{2022arXiv220108666L}, while the tension remains highly significant, its severity has somewhat lessened in the past two years. Many recent near-field observations yield similar results as the CMB measurements for $H_0$, including one using Type Ia supernovae but with a different calibration. The hope that the Hubble tension will be resolved in the coming years is reasonable.}. Even if the tension between far- and near-field observations remains, the precision of the near-field measurements will be sufficient to narrow down the prior for $H_0$ to the degree that the approximation is reasonable and \textit{LISA} can only marginally improve the inference of $H_0$. 

For parallel channel checking, the ground-based GW detectors should already have detected many BNS events to constrain the value of $H_0$ by that time \citep{ECU2023-14032}, so \textit{LISA} could only contribute weakly and $H_0$ could be approximated as fixed again. In principle, BNS and SMBHBs serve in the same way as standard sirens and thus experience similar systematics. However, SMBHBs are much more luminous than BNS and thus detectable up to higher redshift with stronger power in constraining $\Omega_m$ and $\Omega_{\Lambda}$. Therefore, we focus on inferring $\Omega_m$ and $\Omega_{\Lambda}$ in this paper. 

In this paper we also do not perform joint cosmological inference. Therefore, we may write $p(\Vec{\Omega}|\, \mathbf{z}_{\rm s}\,,\mathbf{d}_{\rm lens}) = p(\Vec{\Omega})$ and we assume for simplicity flat priors on the cosmological parameters. The posterior on cosmological parameters then becomes
\begin{equation}
p(\Vec{\Omega}|\,\mathbf{D}_{\rm GW}\,, \mathbf{z}_{\rm s}\,,\mathbf{d}_{\rm lens}) \propto p(\mathbf{D}_{\rm GW}\,|\,\Vec{\Omega}, \mathbf{z}_{\rm s}\,,\mathbf{d}_{\rm lens}) \,.
\end{equation}
The $\mathbf{D}_{\rm GW}$ term includes the observed luminosity distance $D_L^{\rm obs}$ and their measurement uncertainties $\sigma_{D_L}$, where the latter are obtained from the Fisher matrix formalism with first-order approximation \citep{zhuConstrainingCosmologicalParameters2022a, chassande-mottinGravitationalWaveObservations2019}. Therefore, the likelihood $p(\mathbf{D}_{\rm GW}\,|\,\Vec{\Omega}, \mathbf{z}_{\rm s}\,,\mathbf{d}_{\rm lens})$ can be derived as\footnote{See Appendix~\ref{appd} for a detailed derivation.}
\begin{equation}
\begin{split}
    p(\mathbf{D}_{\rm GW}|\,\Vec{\Omega}, \mathbf{z}_{\rm s},\mathbf{d}_{\rm lens}) \propto \\
    \prod_i \int & \exp\left[D_L^{\rm cor}(\Vec{\Omega}, z_{\rm s}, \mu - \mu_{\rm est}) - D_L^{\rm obs, cor})^2/2\sigma_{D_L}^2 \right] \\ 
    &  \times p(\mu - \mu_{\rm est}|\,\mathbf{d}_{\rm lens}, z_{\rm s})\, d\mu \,,
\end{split}
\end{equation}
where $\mu_{\rm est}$ is the estimated magnification, $\mu$ is the true magnification, and $z_{\rm s}$ is the redshift of each siren. The corrected luminosity distances $D_L^{\rm cor}(\Vec{\Omega}, z_{\rm s}, \mu - \mu_{\rm est})$ and $D_L^{\rm obs, cor}$ are determined by,
\begin{equation}
\begin{split}
&D_L^{\rm cor}(\Vec{\Omega}, z_{\rm s}, \mu - \mu_{\rm est}) = D_L(\Vec{\Omega}, z_{\rm s}) \times (1 + (\mu_{\rm est} -\mu)/2)\,,\\
&D_L^{\rm obs, cor} = D_L^{\rm obs}\sqrt{\mu_{\rm est}}
\end{split}
\end{equation}
where $D_L(\Vec{\Omega}, z_{\rm s})$ is the luminosity distance calculated by Eq.~\eqref{lum_z}.

The $p(\mu - \mu_{\rm est}|\,\mathbf{d}_{\rm lens},z_{\rm s})$ term is the conditional distribution of the error in magnification estimation $\mu - \mu_{\rm est}$ given the lensing data $\mathbf{d}_{\rm lens}$ at redshift $z_{\rm s}$. The construction of $\mu_{\rm est}$ and the estimation of $p(\mu - \mu_{\rm est}|\,\mathbf{d}_{\rm lens}, z_{\rm s})$ is given in Sec.~\ref{estimator}.

\subsection{Weak-lensing reconstruction simulation}\label{sec:wek_recon}
In Section~\ref{sec:wl&delensing}, it was noted that the weak-lensing error is connected with the observed luminosity distance $D^{\rm obs}_L$ by means of the magnification $\mu$. The magnification is further related to the lensing convergence $\kappa$ by \citep{Takahashi_2006,shapiroDelensingGravitationalWave2010b},
\begin{equation}\label{eq:mu}
\mu\approx1+2\kappa.
\end{equation}
Therefore, if we can resolve the convergence by constructing an accurate convergence map around the siren’s location, then we can estimate the siren's weak-lensing error. The convergence map can be constructed from other weak-lensing fields, which is a method called weak-lensing reconstruction. Here we only focus on the use of weak-lensing shear fields, leaving a discussion about using other lensing fields like flexion until Sec.~\ref{Discussion}.

In this paper, we make use of the simulated weak-lensing maps from the extended Scinet Light Cone Simulations (SLICS) \citep{harnois-derapsCosmologicalSimulationsCombinedProbe2018}. The SLICS contain flat sky weak-lensing maps constructed by ray-tracing with the Multiple-Lens-Plane technique \citep{valeSimulatingWeakLensing2003a} under the Born approximation \citep{schneiderNewMeasureCosmic1998,whiteSimulationsWeakGravitational2004}. It has been shown that these approximations are in good agreement with the full treatment and only deviate slightly at the smallest scale \citep{harnois-derapsSimulationsWeakGravitational2015a,hilbertAccuracyWeakLensing2020a}. The lensing maps neglect baryonic effects and consider lensing by dark matter only. A brief introduction to the Born approximation and Multiple-Lens-Plane technique, together with the reconstruction algorithm, is as follows.

\subsubsection{Ray-tracing}
The trajectories of photons are deflected gravitationally by intervening matter inhomogeneities before they reach the observer. This deflection is called gravitational lensing, which causes position shifts in the observed image \citep{schneiderGravitationalLensingStrong2006}. Therefore, the weak-lensing simulations are generally constructed by integrating over null geodesics to calculate the total deflection along the past light cone \citep{harnois-derapsSimulationsWeakGravitational2015a}. Under the Born approximation, the integrations are simplified to be calculated on straight lines (rather than photons' trajectories). Then the weak-lensing convergence $\kappa(\boldsymbol{\theta})$ can be obtained by integrating over the density contrast $\delta(\boldsymbol{\theta},\chi)$ along the line of sight:
\begin{equation} \label{eq:4}
\kappa(\boldsymbol{\theta},\chi_{s}) = \frac{3H^2_0\Omega_m}{2c^2} \int^{\chi_{s}}_0 \frac{\chi(\chi_s -\chi)}{\chi_s}\delta(\boldsymbol{\theta},\chi)(1+z) \, d\chi,
\end{equation}
where $\chi$ is the comoving distance and $\chi_s$ is the comoving distance to the source plane in which the targeted siren is embedded. The density contrast $\delta(\boldsymbol{\theta},\chi)$ is defined by
\begin{equation} \label{eq:5}
\delta(\boldsymbol{\theta},\chi) \equiv \frac{\rho(\boldsymbol{\theta},\chi)}{\bar{\rho}(\chi)} - 1 \,,
\end{equation}
where $\rho$ is the matter density and $\bar{\rho} = 3H^2_0 \Omega_m/8\pi Ga^3$ is the mean density of the universe at different time. $H_0$ and $\omega_m$ are the present-day values of the Hubble constant and density parameter. 

Furthermore, the matter distribution in the light cone can be approximated as a set of discrete lens planes. Then the 3D matter density distribution can be collapsed into planes of two-dimensional density fluctuations, given by,
\begin{equation} \label{eq:2D_density_contrast}
\delta_{2D}(\boldsymbol{\theta},\chi_{\rm lens}) = \frac{1}{\Delta\chi}\int^{\chi_{\rm lens} + \frac{1}{2}\Delta\chi}_{\chi_{\rm lens} - \frac{1}{2}\Delta\chi} \delta(\boldsymbol{\theta},\chi) \, d\chi,
\end{equation}
where $\Delta\chi$ denotes the comoving length of the collapsed region. This effectively turns the integration along the light ray into a discrete sum at the lens locations and manifests the name of the technique. The convergence $\kappa$ is then computed by,
\begin{equation} \label{eq:7}
\kappa(\boldsymbol{\theta},\chi_{s}) = \frac{3H^2_0\Omega_m}{2c^2} \sum^{\chi = \chi_n}_{\chi = \chi_1} \frac{\chi(\chi_s -\chi)}{\chi_s}\delta_{2D}(\boldsymbol{\theta},\chi)(1+z(\chi))\Delta\chi,
\end{equation}
where $\chi_1$ is the comoving distance to the first lens plane and $\chi_n$ is the comoving distance to the last lens plane before the source.

\subsubsection{KS inversion method}
The weak-lensing shear and convergence are both combinations of derivatives of a scalar lensing potential field $\psi(\boldsymbol{\theta})$. The two components of the shear can be written in terms of $\psi(\boldsymbol{\theta})$ by,
\begin{equation} \label{eq:2}
\gamma_{1} = \frac{1}{2}(\partial^2_{1} -\partial^2_{2})\psi,\,\,\,
\gamma_{2} = \partial_{1}\partial_{2}\psi \,,
\end{equation}
where the partial derivatives $\partial_{i}$ with $i = 1, 2$ correspond to the angular coordinates $\theta_{i}$. The convergence $\kappa(\boldsymbol{\theta})$ can also be expressed in terms of $\psi(\boldsymbol{\theta})$ by
\begin{equation} \label{eq:3}
\kappa = \frac{1}{2}(\partial^2_{1} + \partial^2_{2})\psi \,.
\end{equation}
Then the shear maps $\gamma_{1,2}(\boldsymbol{\theta})$ can be computed via Fourier transformation by,
\begin{equation} \label{eq:gamma_comp}
\hat{\gamma}_{1,2} = \hat{\kappa}\,\hat{P}_{1,2}\,,
\end{equation}
where the hat symbol denotes Fourier transform and $\hat{P_{1}}(\boldsymbol{\ell}), \hat{P_{2}}(\boldsymbol{\ell})$ are given by,
\begin{equation} \label{eq:9}
\hat{P_{1}}(\boldsymbol{\ell}) = \frac{\ell^2_1 - \ell^2_2}{\ell^2},\,\,\,
\hat{P_{2}}(\boldsymbol{\ell}) = \frac{2\ell_1\ell_2}{\ell^2},
\end{equation}
with $\ell^2 \equiv \ell^2_1 + \ell^2_2 $ and $\ell_i$ being the wave numbers with respect to the angular coordinates $\theta_{i}$.

The order is reversed in practice as the convergence map is not directly observable in a galaxy survey. By contrast, the shear maps $\gamma_{1,2}(\boldsymbol{\theta}, z)$ can be derived from the weighted average ellipticities of the galaxies around the position $\boldsymbol{\theta}$ in the image\footnote{The ellipticity of a galaxy is a point estimate for the reduced shear at that sky position and redshift, which is the only real observable in galaxy surveys. We ignore the reduced factor as it is close to one in the weak-lensing regime.}. The weighting takes factors like the redshift distribution of galaxies into consideration and more details will be provided in the second half of Sec.~\ref{perf}. Then the KS inversion method \citep{1993ApJ...404..441K,bartelmannWeakGravitationalLensing2001a} can be used to reconstruct the convergence map\footnote{If there is no available information about some line-of-sights (due to sparsity of galaxies or masks), then advanced methods, such as the one in \citet{piresEuclidReconstructionWeaklensing2020}, should be applied to overcome the missing-data problem.}. The KS inversion method is simply the inverse of Eq.~\eqref{eq:gamma_comp}, 
\begin{equation} \label{eq:k_space_ks}
\hat{\kappa} = \hat{P_{1}}\hat{\gamma_{1}} + \hat{P_{2}}\hat{\gamma_{2}} \,,
\end{equation}
where the hat symbol denotes Fourier transforms and $\hat{P_{1}}(\boldsymbol{\ell}), \hat{P_{2}}(\boldsymbol{\ell})$ are the same as above. According to Eq.~\eqref{eq:k_space_ks}, the shear maps $\gamma_{1,2}(\boldsymbol{\theta})$ are related to the convergence map $\kappa(\boldsymbol{\theta})$ by simple algebraic equations in the Fourier domain. Then the reconstructed convergence map can be obtained by inverse Fourier transforms.

\subsubsection{Mass-sheet degeneracy problem}\label{mass_sheet}
There is a degeneracy when reconstructing $\kappa$ from $\gamma_{1,2}$ when $\ell_1 = \ell_2 = 0$. Consequently, the mean value of the reconstructed convergence field $\Bar{\kappa}$ cannot be determined only from shear information, which is
the so-called mass-sheet degeneracy \citep{1995A&A...303..643B}. In practice, the observational area is finite, resulting in a lower bound $\ell_{\rm min}$ for the wave numbers of the reconstructed Fourier modes. Basically, the reconstructed convergence would differ from the true convergence field by a constant determined by fluctuations larger than the field area.

For wide-field surveys, we often set the reconstructed modes with $\ell \leq \ell_{\rm min}$ to zero, as the fluctuations larger than the field area are negligible. This is a reasonable assumption for wide-field reconstruction \citep{masseyDarkMatterMaps2007}. For deep-field surveys, the observation area is much smaller and thus the error from mass-sheet degeneracy must be treated seriously.

In principle, the mass-sheet degeneracy problem for deep-field surveys may be alleviated if additional wide-field surveys around that region are available \citep{shapiroDelensingGravitationalWave2010b}. The basic idea behind this is that the deep images will be used to measure small-scale convergence fluctuations, while the wide images will pick up modes larger than the size of the deep images. In practice, the effectiveness of hybridizing wide and deep survey maps is subject to weak-lensing shape noise, which highly depends on the size of the deep-field survey as well as the galaxy density $n_{\rm gal}(\boldsymbol{\theta}, z_{\rm gal})$ of the wide-field survey. It is possible that the hybridization has no improvement because the galaxy density of the wide-field survey is not enough to obtain a satisfying signal-to-noise ratio (SNR) even in probing large-scale fluctuations.

\subsubsection{Weak-lensing shape noise}
The gravitational shears can be derived from the ellipticities $\epsilon_{1,2}$ of the background galaxies. However, in reality, the projected shapes of galaxies are not intrinsically circular. The measured ellipticity is a combination of their intrinsic ellipticity and the gravitational lensing shear. The shear is also subject to measurement noise and uncertainties in the PSF (point spread function) correction. All these effects can be modelled as additive noises\footnote{In reality, the measurement noise and uncertainties in the PSF correction are much more complicated and highly detector dependent. The effects should be a combination of additive or multiplicative errors/biases. Here,  we are evaluating the potential of delensing under very ideal conditions.} to both components of the shear field \citep{piresEuclidReconstructionWeaklensing2020}, 
\begin{equation} \label{eq:10}
\epsilon_{i} = \gamma_{i} + N_{i},
\end{equation}
where $i = 1,2$. The noises $N_{1}$ and $N_2$ are assumed to be Gaussian and independent of each other, with zero mean and standard deviation given by,
\begin{equation} \label{eq:11}
\sigma^{i}_{s}(\boldsymbol{\theta}) = \frac{\sigma_{\gamma}}{\sqrt{n_g(\boldsymbol{\theta})}},
\end{equation}
where $n_{\rm gal}(\boldsymbol{\theta})$ is the number of galaxies within the pixel at $\boldsymbol{\theta}$. The shear dispersion per galaxy, $\sigma_{\gamma}$, arises both from the measurement uncertainties and the intrinsic scatter of galaxy shapes. The Gaussian assumption is reasonable \citep{hirataReducingWeakLensing2010a} in the weak-lensing regime and here we further assume that the shear dispersion for each galaxy is independent of each other. In other words, we ignore the intrinsic alignment of neighbour galaxies which can bias the shear estimate. 
In light of Eq.~\eqref{eq:10}, we derive in Fourier space,
\begin{equation} \label{eq:12}
\hat{\kappa}^{\,\rm noisy} = \hat{\kappa} + \hat{P_{1}}\hat{N_1} + \hat{P_{2}}\hat{N_2}\,.
\end{equation}
Then the reconstructed convergence map is also subject to an additive noise $N_{\kappa}$ by,
\begin{equation} \label{eq:13}
N_{\kappa} = {P_{1}}\ast{N_1} + {P_{2}}\ast{N_2} \,,
\end{equation}
where the asterisk denotes convolution, $P_1$ and $P_2$ are the inverse Fourier transforms of $\hat{P_1}$ and $\hat{P_2}$.

When the shear noises $N_1$ and $N_2$ are Gaussian and independent across the field with a constant standard deviation, then the induced noise in convergence maps is also Gaussian and independent at each position, with a standard deviation $\sigma_{\kappa} = \sigma^{1}_{s}= \sigma^{2}_{s}$ according to Eq.~\eqref{eq:12}. In reality, the number of galaxies varies slightly across the field. 
Even at the same position, the variances of $N_1$ and $N_2$ might also be slightly different. These effects introduce noise correlations and variations in the reconstructed convergence maps, but they were found to remain negligible compared to other effects studied in this paper. 

The pixelation of the observed ellipticity fields already smooths the fields with a smoothing scale equal to the pixel size. However, it may be insufficient to obtain a satisfactory signal-to-noise ratio for the estimated convergence. Then the raw estimated convergence should be further smoothed by a Gaussian filter with filter scale $\theta_s$, 
\begin{equation} \label{eq:sm_filter}
\kappa_{\rm smooth}(\boldsymbol{\theta}) \propto \int exp(-\frac{(\boldsymbol{\theta}-\boldsymbol{\theta}')^2}{2\theta^2_s})\kappa_{\rm raw}(\boldsymbol{\theta}') \, d^2\boldsymbol{\theta}'
\end{equation}
where the integration is conducted on the whole field area and the proportionality constant is determined by normalization. The smoothing can also be done by using Wiener filters, but it turns out that the improvements are marginal \citep{10.1111/j.1365-2966.2010.17963.x}. 
There should exist an optimal smoothing scale determined by a trade-off between reconstructing small-scale features of the convergence and reducing shape noise by averaging over galaxy images \citep{dalalCorrectiveLensesHighRedshift2003}.

\subsubsection{Redshift distribution of galaxies}
Finally, the galaxies do not all lie at the redshift $z_{\rm s}$ of the standard candle/siren but have a certain redshift distribution $n_{\rm gal}(\boldsymbol{\theta}, z_{\rm gal})$ that depends on their intrinsic redshift distribution and on the depth of the survey. A smoothed version of the effective convergence will be reconstructed if the individual galaxy redshifts are not utilized:
\begin{equation} \label{eq:red_s}
\kappa_{\rm eff}(\boldsymbol{\theta}) = \int n_{\rm gal}(\boldsymbol{\theta}, z_{\rm gal})\, \kappa(\boldsymbol{\theta}, z_{\rm gal})\,dz_{\rm gal}.
\end{equation}

If the redshift distribution of the galaxy number density  $n_{\rm gal}(\boldsymbol{\theta},z_{\rm gal})$ is not sharply peaked around the redshift $z_{\rm s}$ of the standard siren, the effective convergence $\kappa_{\rm eff}(\boldsymbol{\theta})$ may deviate substantially from the true convergence $\kappa(\boldsymbol{\theta}, z_{\rm s})$. Variations in the redshifts of galaxies contribute to an additional noise source.

\subsection{Estimator of magnification from noisy convergence maps} \label{estimator}
The reconstructed convergence maps from shear measurements can be converted directly into the estimated magnification maps simply by Eq.~\eqref{eq:mu}.\footnote{Here we assume the weak-lensing limit, where the lensing convergence, shear and rotation are small.}

However, as pointed out by \citet{10.1111/j.1365-2966.2010.17963.x}, this simple estimate might fail if the estimated convergence $\kappa_{\rm est}(\boldsymbol{\theta})$ deviates substantially from the true convergence $\kappa(\boldsymbol{\theta}, z_{\rm s})$. We quantify this deviation by the residual magnification, 
\begin{equation} \label{residual}
\mu_{\rm res} = \mu - \mu_{\rm est}. 
\end{equation}
The difference may originate from the weak-lensing shape noise, mass-sheet degeneracy, or realistic redshift distribution of galaxies. These effects make the simple estimate perform even worse than just using the lensing prior in certain cases (i.e., assuming $\mu_{\rm est}(\boldsymbol{\theta}) = 0$).

To improve this, one should construct an unbiased magnification estimator which minimizes the residual dispersion $\sigma_{\mu_{\rm res}}$ \citep{10.1111/j.1365-2966.2010.17963.x}. If the conditional distribution $p(\mu|\kappa_{\rm est})$ of the true magnification $\mu$ for a given estimated convergence $\kappa_{\rm est}$ is known, then the estimator satisfying the above requirements can be derived as,
\begin{equation}
\mu_{\rm est}(\boldsymbol{\theta}) \,=\,\, \left\langle\mu \right\rangle_{\mu|\kappa_{\rm est}}(\kappa_{\rm est}(\boldsymbol{\theta})),
\end{equation}
where $\left\langle\mu \right\rangle_{\mu|\kappa_{\rm est}}(\kappa_{\rm est})$ is the expectation value of the true magnification $\mu$ for a given estimated convergence $\kappa_{\rm est}$:
\begin{equation}
\left\langle\mu \right\rangle_{\mu|\kappa_{\rm est}}(\kappa_{\rm est}) = \int^{\mu_{\rm max}}_0 \, \mu \, p(\mu|\kappa_{\rm est})\, d\mu, 
\end{equation}
where $\mu_{\rm max}$ is the maximal allowed value of the magnification which remains in the weak-lensing regime. In the subsequent discussions and analyses, $\mu_{\rm max} = 1.5$ is chosen so that all the data above that value are abandoned.

The conditional distribution of the true magnification given the estimated convergence can be inferred from numerical simulations, thus allowing the optimal magnification estimator to be constructed. However, the magnification estimator constructed in this way will be dependent on the values of the cosmological parameters (and other settings, e.g. the galaxy biasing scheme) adopted by the numerical simulations that are used. Strictly, therefore, this magnification estimator ought to account for uncertainties in these model parameters and settings through appropriate marginalisation -- with the result that, in practice, the optimal magnification estimator would likely perform somewhat more poorly than considered here. In what follows, however, we will not consider this issue further since our purpose in this work is to investigate the limitations of delensing methods in a realistic setting but under more favourable conditions.

In Appendix A of \citet{10.1111/j.1365-2966.2010.17963.x} it is shown that this estimator is optimal in the sense that no other magnification estimator based on the estimated convergence $\kappa_{\rm est}$ yields a smaller dispersion in the residual magnification. For instance, the residual magnification dispersion $\sigma_{\mu_{\rm res}}$ for the best estimator is never larger than the dispersion in the uncorrected case. 

The conditional distributions $p(\mu_{\rm res}|\kappa_{\rm est})$\footnote{$p(\mu_{\rm res}|\kappa_{\rm est})$ and $p(\mu|\kappa_{\rm est})$ only deviate in their mean and have the same standard deviation $\sigma_{\mu_{\rm res}|\kappa_{\rm est}}=\sigma_{\mu|\kappa_{\rm est}}$ by construction.} characterizes the accuracy of the weak-lensing reconstruction and is equivalent to $p(\mu - \mu_{\rm est}|\,\mathbf{d}_{\rm lens}, z_{\rm s})$ in Sec.~\ref{baye}. The redshift dependence does not show explicitly in $p(\mu_{\rm res}|\kappa_{\rm est})$ for simplicity of notation. To evaluate the general delensing performance, $p(\mu_{\rm res}|\kappa_{\rm est})$ could be marginalized to obtain the distribution $p(\mu_{\rm res})$ for the specific observation scheme at a particular redshift. 

Although the lensing distributions like $p(\mu_{\rm res})$ and $p(\mu)$ are in principle non-Gaussian \citep{hirataReducingWeakLensing2010a,2011MNRAS.411....9S}, we approximate them as Gaussian and only focus on the standard deviations $\sigma_{\mu_{\rm res}}$ when discussing the weak-lensing error and delensing performance.

\subsection{Simulations}\label{sim}
\subsubsection{SLICS Catalogs}\label{SLICS_intro}

The SLICS (Scinet LIght Cone Simulations) \citep{harnois-derapsSimulationsWeakGravitational2015a} were designed as a massive upgrade of the CLONE simulations \citep{harnois-derapsGravitationalLensingSimulations2012a}. 
In our work, we used the expanded version of the SLICS suite \citep{harnois-derapsCosmologicalSimulationsCombinedProbe2018}, including hybrid mock catalogs that represent future lensing data at the level of LSST (\textit{Large Synoptic Survey Telescope}, now designated the Vera Rubin Observatory) \citep{2019ApJ...873..111I}. 
The most significant advantage of the LSST-like hybrid mock catalogs is that they simultaneously contain lensing and galactic properties. 

The SLICS are based on 1025 N-body simulations produced by the high-performance gravity solver {\small CUBEP3M} \citep{harnois-derapsHighperformanceP3MNbody2013a}. 
The series of N-body simulations were used to construct dark matter halo catalogs, which later served as the galaxy catalogs' skeleton. Then mock galaxy catalogs are produced from Halo Occupation Distribution (HOD) models with only mass dependence. The luminosity of the whole galaxy in the \textit{r}-band is also given for each galaxy in the catalogs, consistent with the HOD used.

Throughout this work, we neglect the higher-order correction caused by wave optics lensing\footnote{Wave optics lensing refers to the suppression of magnification when the gravitational-wave wavelength matches approximately the Schwarzschild radius of the gravitational lens.}. The reason is that the mass resolution of SLICS is not adequate to resolve the matter fluctuations with scales sensitive to the wave nature of GWs at \textit{LISA} frequencies~\citep{Takahashi_2006,Oguri:2020ldf,10.1093/mnras/stv794}. In reality, the wave optics effect would slightly suppress the lensing magnification experienced by the GWs detected by \textit{LISA}. 

The same set of N-body simulations of dark matter was used to construct shear and convergence maps via the ray-tracing algorithm with the multiple-plane tiling technique. The maps were all flat-sky, $100\;{\rm deg}^{2}$ maps with $7745^2$ pixels, computed on specified lens planes. The redshifts of the lens planes $z_{\rm lens}$ and source planes $z_{\rm source}$ are listed in Table~\ref{tab:z_planes}.
Once the galaxy catalogs are given, the lensing information can be linearly interpolated at the galaxy coordinates and redshifts from the lens planes.
The interpolation is only done along the redshift direction since the resolution of the lens maps already approaches the limitation in the mass resolution of the N-body simulations. 
In addition to the shear, the observed ellipticity is also included in the LSST-like hybrid catalogs. 
The observed ellipticity $\epsilon_{1,2}$ deviates from the true shear $\gamma_{1,2}$ by a Gaussian error with width $\sigma_\gamma = 0.29$ per galaxy, where the error takes both the intrinsic shape noise and measurement errors into account. 

\begin{table*}
    \centering
    \begin{tabular}{p{0.05\linewidth} p{0.025\linewidth} p{0.025\linewidth} p{0.025\linewidth} p{0.025\linewidth} p{0.025\linewidth} p{0.025\linewidth} p{0.025\linewidth} p{0.025\linewidth} p{0.025\linewidth} p{0.025\linewidth} p{0.025\linewidth} p{0.025\linewidth} p{0.025\linewidth} p{0.025\linewidth} p{0.025\linewidth} p{0.025\linewidth} p{0.025\linewidth} p{0.025\linewidth}}
    \hline
    $z_{\rm lens}$ & 0.042 & 0.130 & 0.221&  0.317&  0.418 & 0.525 & 0.640 & 0.764 & 0.897 & 1.041 & 1.199 & 1.373
    & 1.562 & 1.772 & 2.007 & 2.269 & 2.565 & 2.899\\
    \hline
    $z_{\rm source}$ & 0.086 & 0.175 & 0.268 & 0.366 & 0.471 & 0.582 & 0.701 & 0.829 & 0.968 & 1.118 & 1.283 & 1.464 & 1.664 & 1.886 & 2.134 & 2.412 & 2.727 & 3.084\\
    \hline   
    \end{tabular}
    \caption{The redshift of lens and source planes used for ray-tracing with the Multiple-Lens-Plane technique to construct the weak-lensing maps.}
    \label{tab:z_planes}
\end{table*}

\begin{table*}
    \centering
    \begin{tabular}{c|c|c|c|c}
    \hline
    \multicolumn{5}{c}{LSST-like hybrid catalogs contain information about each galaxy:} \\
    \hline
    \multicolumn{5}{c}{sky position $x,y$ ; redshift $z$ ; Luminosity in \textit{r}-band $L_{r}$ ; convergence $\kappa$ ; shear $\gamma_{1,2}$ ; observed ellipticity $\epsilon_{1,2}$}\\
    \hline\hline
    Total area of the field & The redshift range &	Galaxy number density & Resolution of lensing maps & Shear dispersion (single galaxy)\\
    \hline
    $10 \times 10 \,{\rm deg}^2$ & $0\,-\,3$ & $28.3\,{\rm arcmin}^{-2}$ & $\approx\ 4.6\:{\rm arcsec}$ & $\sigma_\gamma = 0.29$\\
    \hline
    
    \end{tabular}
    \caption{Relevant information about the LSST-like hybrid catalogs.}
    \label{tab:LSST}
\end{table*}

All the information about the LSST-like hybrid catalogs (LSST-like catalogs hereafter) related to our work is summarized in Table~\ref{tab:LSST}. For details of the SLICS catalogs, please refer to the extended SLICS paper \citep{harnois-derapsCosmologicalSimulationsCombinedProbe2018}.

\subsubsection{Siren Catalogs}\label{GW_sim:sec} 

Catalogs of standard sirens are necessary to emulate the realistic delensing outcomes. The construction of these catalogs depends on our understanding of the universe, specifically on the SMBHB merger models. Apart from modelling the sources, the observed GW signals vary with the detector configurations. In this paper, we used the siren catalogs generated by \citet{zhuConstrainingCosmologicalParameters2022a}, which consider all the factors mentioned above and include all the ingredients needed for our purpose.

The siren catalogs in \citet{zhuConstrainingCosmologicalParameters2022a} began with catalogs of SMBHB mergers. These preliminary catalogs depend on possible formation models of supermassive black holes (SMBHs) including ‘popIII’, ‘Q3d’ and ‘Q3nod’. Basically, the redshift and mass distributions as well as the average number of mergers depend on the formation models. An introduction to the formation models can be found in \citet{zhuConstrainingCosmologicalParameters2022a} and they are not our main focus in this paper. Note that for both ‘popIII’, ‘Q3d’ and ‘Q3nod’, the evolution of the SMBHs should be deeply correlated with the evolution of their host galaxies \citep{ferrareseFundamentalRelationSupermassive2000,dingMassRelationsSupermassive2020a}. 

Then the GW waveforms were generated from the IMRPhenomPv2 \citep{hannamSimpleModelComplete2014a} according to the merger catalogs. After the waveform generation, the siren catalogs were constructed based on the waveforms and multiple configurations for the space-borne GW detectors. The detector configurations included individuals and combinations of potential space-borne GW detectors. Here we concentrate on the detector configuration \textit{TQ}+\textit{LISA}. Only sirens with redshift $z_s<3$ and producing GW signals with SNR $\rho>8$ were presented in the siren catalogs\footnote{For signals with SNR just below the threshold, lensing might help the signals to be detected by slightly magnifying their waveforms. However, the probability of such events occurring should be small and the errors in the distance estimation should be large even after being magnified. Therefore, we ignore this issue in our analysis. Notice that this is not the case in the strong-lensing regime as the magnification can be very large so that lensing can significantly increase the SNR of the signals.}.

Each realization of the siren catalogs includes the following information about each siren: the true luminosity distance $D_{L}$, the measurement error of the luminosity distance $\sigma_{D_{L}}$, the source redshift $z_s$, and the total mass of the SMBHB $M_{\rm tot}$. For more details about the siren catalogs, please refer to \citet{zhuConstrainingCosmologicalParameters2022a}. 
    
\subsubsection{Combining two simulations}\label{comb}
In Section~\ref{GW_sim:sec}, we pointed out that the evolution of the SMBHs is deeply coupled with the evolution of their host galaxies. Therefore, the siren catalogs should be correlated with the LSST-like galaxy catalogs in a complicated way. However, the catalogs of standard sirens that we are using are based on a statistical description of the formation models and the detector configurations without referring to any specific galaxy catalogs. This freedom enables us to tailor the siren catalogs to the LSST-like hybrid catalogs -- i.e. we can choose host galaxies that are appropriate to the sirens as long as their redshift and mass distributions are consistent.

Here we adopt an empirical relationship reported by \citet{dingMassRelationsSupermassive2020a} to match sirens with their host galaxies. The empirical relationship connects the mass of the central SMBH with the luminosity of its host galaxy in the \textit{r}-band, shown in Eq.~\eqref{eq:empirical}:
\begin{equation} \label{eq:empirical}
\log(\frac{M_{BH}}{10^{7}M_{\odot}})=	0.49+0.90\log(\frac{L_{r}}{10^{10}L_{\odot}})
\end{equation}

In general, there should be a considerable time lag between the merger of galaxies and the merger of their embedded SMBHs \citep{volonteriEvolutionMassiveBlack2007a}. Therefore, we assume that the galaxy has reached its new equilibrium before the inspiraling phase of the SMBHB. Then the properties of the siren are highly correlated to the properties of its host galaxy. Therefore, the above empirical relationship holds with the mass of central SMBH equal to the total mass of the binary. 

We interpret the empirical relationship as a loose constraint with respect to the siren catalogs. We start by binning the galaxies into redshift intervals with bin size $\Delta z=0.1$. Then we implement the empirical relationship to find the most suitable host galaxy for each siren within the redshift bin. 

However, many expected host galaxies of the detected sirens would lie below the luminosity threshold for observation associated with the SLICS catalogues. The sirens that produce GWs detectable by \textit{LISA} and \textit{TQ} have a total mass that is approximately independent of redshift, even in the high redshift region. Thus, the expected luminosity of those host galaxies does not change too much. This presents a problem as we go to higher redshifts, where those galaxies would be too dim to be observed. Furthermore, the empirical relationship might not be valid at high redshift in any case. These issues require future study, and for the present work we simply assign the galaxy with the closest redshift as the host for siren where there is no more suitable choice. We will discuss this problem later in Sec.~\ref{Conclusion}.

The more technical problem is that the two suites of simulations (i.e. those underpinning the SLICS catalogs and our siren catalogs) have slightly different sets of cosmological parameters. From Eq.~\eqref{eq:7} and Eq.~\eqref{eq:10}, it is clear that the weak-lensing maps depend on the cosmological parameter configurations. Since we focus more on the impact of weak gravitational lensing in this paper, for our sirens we adopt a flat $\Lambda CDM$ model with cosmological parameters taken to be the same as in SLICS. As mentioned above, the redshift and mass distributions are the core assumptions for the statistical behaviors of standard sirens. Hence we recompute the luminosity distance of each siren with its redshift $z$ unchanged but with the cosmological parameters re-tuned to match those of the SLICS configurations.

The cosmological parameters adopted in SLICS relied on the best fit $\rm WMAP9 + BAO + SN$ parameters \citep{hinshawNINEYEARWILKINSONMICROWAVE2013}: $\,\mathit{\Omega_{m}}=0.2905,\,\mathit{\Omega_{\Lambda}}=0.7095,\,\mathit{\Omega_{b}}=0.0473,\,h=0.6898, \, \sigma_8 = 0.826$ and $n_s = 0.969$. This choice lies close to the mid-point between the cosmic shear and the Planck best-fit values in the $[\sigma_8- \mathit{\Omega_{m}}]$ plane.

\section{Results}\label{result}
\subsection{Perfect delensing}\label{perf}

We first consider the case of perfect reconstruction, which can be regarded as a hard limit for the more realistic delensing scenarios that we consider later. In this case, the noise-free shear information around the siren at the same redshift $z_s$ is directly accessible, and the information for each line-of-sight is viewed as a sample from the underlying distribution $p(\mu_{\rm res}|\kappa_{\rm est})$. 

\begin{figure}
  \graphicspath{ {./ZF_fig/} }
  \includegraphics[width=\linewidth]{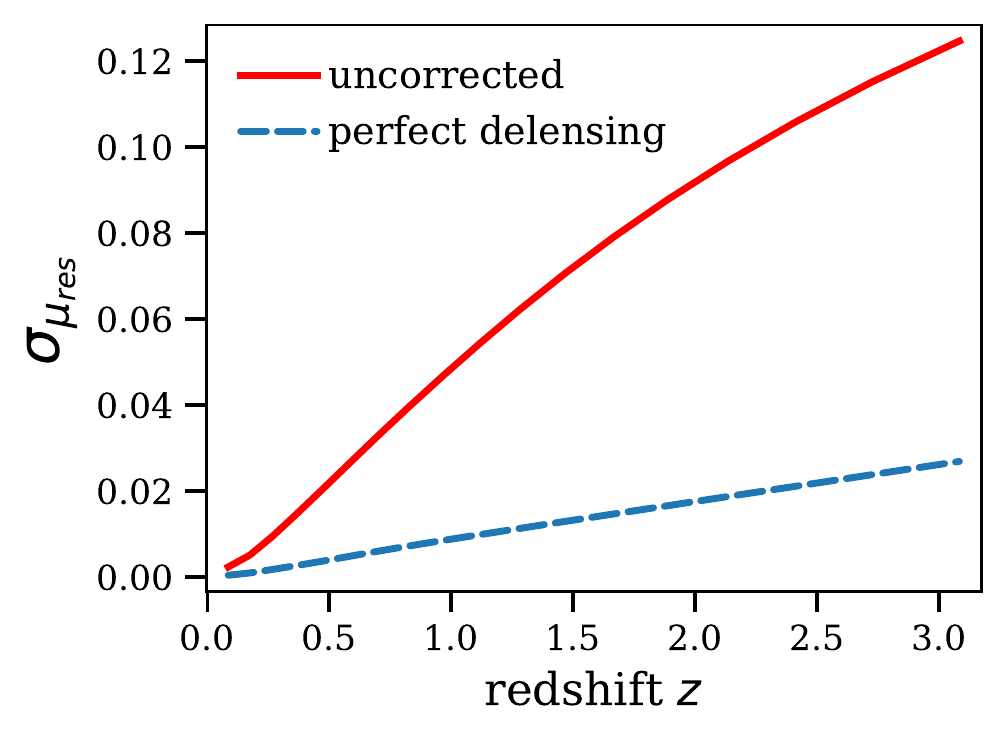}
  \caption{The dispersion of the residual weak-lensing magnification $\sigma_{\mu_{\rm res}}$ for the uncorrected case ($\mu_{\rm est} = 1$, solid line) and the perfectly corrected case as a function of redshift $z$. The perfectly corrected case has a much smaller but non-vanishing standard deviation for the residual $\sigma_{\mu_{\rm res}}$. The small dispersion is due to interpolation error $\sigma_{\mu_{\rm sim}}$ in SLICS.}
  \label{fig:per_err}
\end{figure}

Fig.~\ref{fig:per_err} shows the outcomes of perfect delensing. The uncorrected case ($\mu_{\rm est} = 1$) has a standard deviation $\sigma_{\mu}=0.12$ for standard sirens at redshift $z_s = 2.9$. In contrast, the perfectly corrected case has five times smaller standard deviation for the residual $\sigma_{\mu_{\rm res}}=0.024$. Note that the dispersion of residual magnification in the perfect case is small but non-vanishing. This discrepancy originates from the interpolation setup in SLICS, where the shear maps are constructed by Fast Fourier Transform from the convergence maps, but the interpolation on the pixels is done at the very end, on the shear and convergence maps all at once \citep{harnois-derapsGravitationalLensingSimulations2012a}. Therefore, this small error may be interpreted as the contribution from small-scale fluctuations beneath the resolution of the simulation. 

\begin{figure}
  \graphicspath{ {./ZF_fig/} }
  \includegraphics[width=\linewidth]{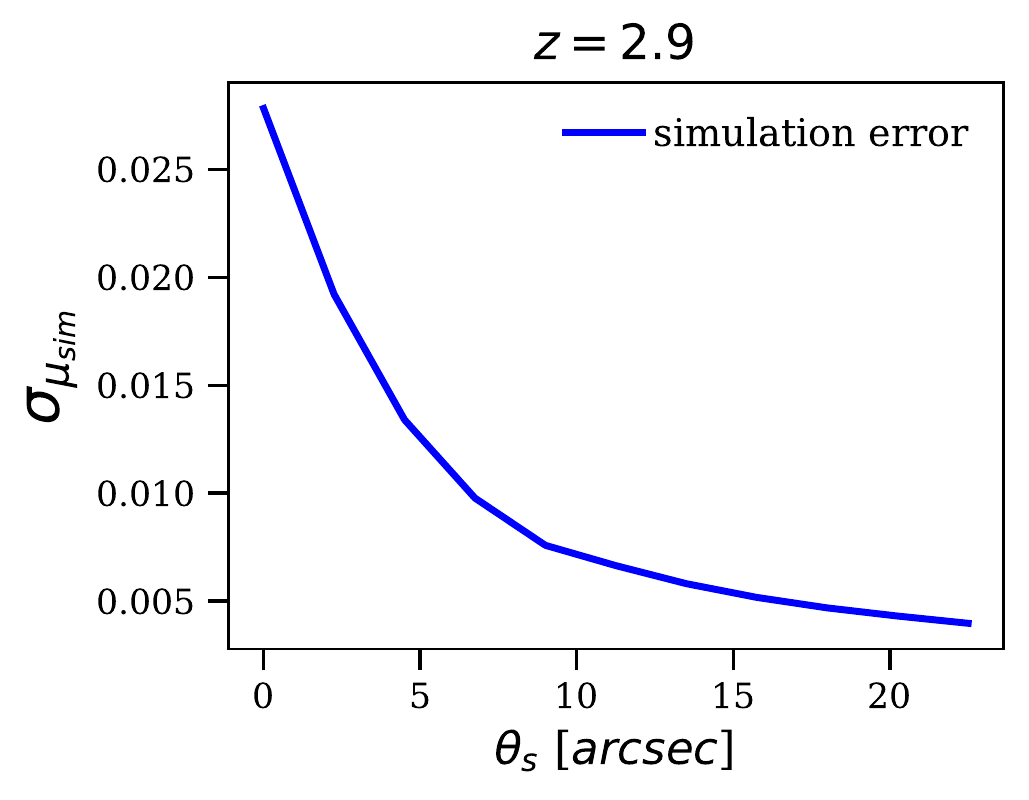}
  \caption{The dispersion due to interpolation uncertainties $\sigma_{\mu_{\rm sim}}$ as a function of the smoothing scale $\theta_s$, for a siren at redshift $z= 2.9$. The $\sigma_{\mu_{\rm sim}}$ decreases rapidly with smoothing and thus can be ignored in realistic cases.}
  \label{fig:per_sm}
\end{figure}

Although the residual error inherited from the numerical simulation, $\sigma_{\mu_{\rm sim}}$, is annoying, Fig~\ref{fig:per_sm} shows that $\sigma_{\mu_{\rm sim}}$ decreases rapidly with smoothing. In realistic cases, smoothing is inevitable to suppress the noise, and hence the interpolation errors are negligible. Therefore, this error will be ignored hereafter unless otherwise specified.
 
\begin{figure}
  \graphicspath{ {./ZF_fig/} }
  \includegraphics[width=\linewidth]{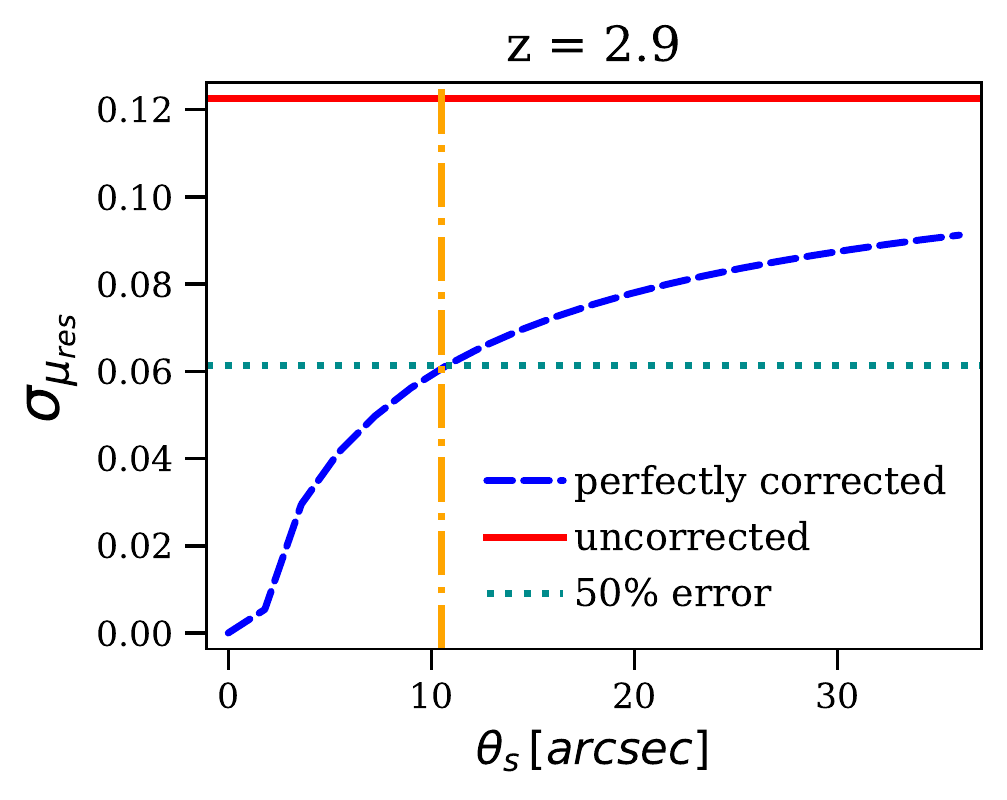}
  \caption{The dispersion of residual magnification $\sigma_{\mu_{\rm res}}$ (dashed line) for sirens at redshift $z_s = 2.9$ as a function of the smoothing filter scale $\theta_s$ from a perfect reconstruction. The smoothing wipes away the small-scale fluctuations and hence the non-vanishing $\sigma_{\mu_{\rm res}}$ is contributed from the modes with scale below $\theta_s$. 
  The solid horizontal line marks the magnification dispersion without correction and
  the smoothing scale corresponding to 50\% of the uncorrected dispersion is denoted in the figure by the dotted horizontal line. The interpolation errors $\sigma_{\mu_{\rm sim}}$ have already been eliminated.}
  \label{fig:per_sm_2.9}
\end{figure}

Even with perfect measurements for individual galaxies, a magnification estimate involves a certain amount of pixelation and smoothing of the shear observation maps, since the inversion algorithm only applies to smooth fields. Also, in realistic cases, the noises in the shape measurements can be reduced by smoothing and averaging. According to Eq.~\eqref{eq:k_space_ks}, the smoothing on the shear measurements eventually propagates into the constructed convergence fields and thus leads to errors in estimating the magnification. The impact of smoothing on the dispersion of the residual magnification $\sigma_{\mu_{\rm res}}$ at $z = 2.9$ is shown in Fig.~\ref{fig:per_sm_2.9}, where the interpolation error has already been neglected. As shown in Fig.~\ref{fig:per_sm_2.9}, the magnification correction is already degraded substantially (reaching 50\% of the uncorrected dispersion) if the convergence map is smoothed with a filter scale\footnote{The definition of smoothing filter scale $\theta_s$ is given in Eq.~\eqref{eq:sm_filter}.} $\theta_s\approx10$ arcsec. This illustrates that it is necessary to reconstruct the convergence on scales of a few arcseconds in order to reduce the lensing error by a factor of two or better, compared to the uncorrected case. 

The perfect case also helps to reduce the error caused by the realistic redshift distribution of galaxies. As pointed out in Eq.~\eqref{eq:red_s}, if the redshift information of the galaxies is not available, only an effective convergence $\kappa_{\rm eff}(\boldsymbol{\theta})$ can be deduced from the realistic redshift distribution of galaxies, which may deviate substantially from the actual convergence at the sirens $\kappa(\boldsymbol{\theta}, z_{\rm s})$. This error further propagates into the estimation of magnification. If the redshift information, either photometric or spectroscopic, about the galaxies is accessible, the variations of the lensing properties as a function of redshift can be exploited to perform a three-dimensional reconstruction of the estimated magnification maps. However, the correlation across lens planes has been explicitly broken due to the way of construction in SLICS, hence any 3D reconstruction should be performed only inside individual lens sub-volumes. Here, we take a simpler approach that is similar to \citet{10.1111/j.1365-2966.2010.17963.x}, which does not attempt to construct the 3D matter structures causing the lensing along the line of sight but only exploits the statistical relation between lensing quantities at different discrete redshifts. 

\begin{figure}
  \graphicspath{ {./ZF_fig/} }
  \includegraphics[width=\linewidth]{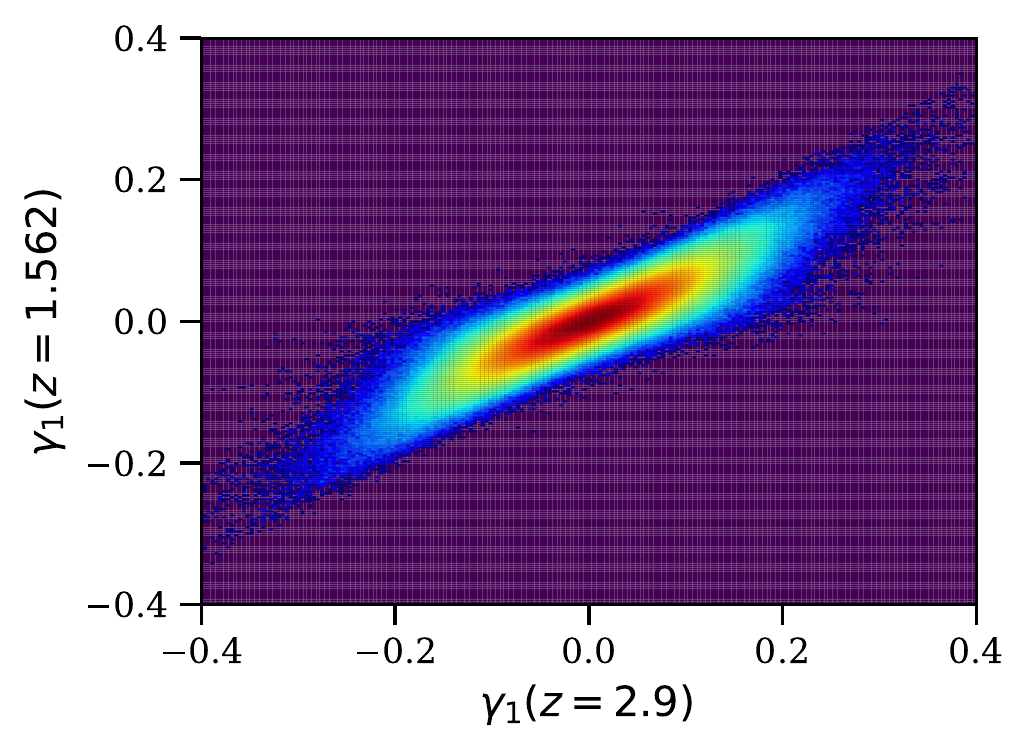}
  \caption{The joint distribution of the shear component $\gamma_1$ at redshift $z = 1.562$ and redshift $z_s = 2.9$,  at the same sky position. Lighter areas correspond to higher probability densities on a logarithmic scale. The joint distribution of $\gamma$ at different redshifts along the same line of sight is highly correlated and well approximated by a linear relation. Note that the joint distribution for the second shear components $\gamma_2(z)$ and $\gamma_2(z_s)$ is the same as for the first shear components.}
  \label{fig:correlation}
\end{figure}

\begin{figure}
  \graphicspath{ {./ZF_fig/} }
  \includegraphics[width=\linewidth]{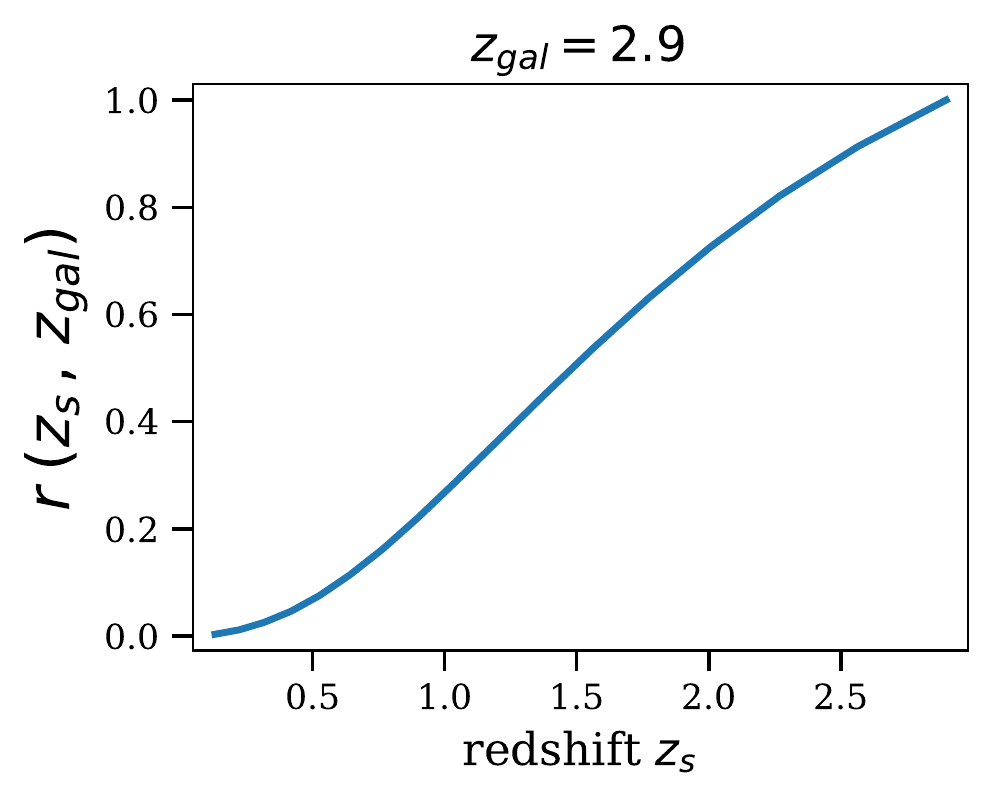}
  \caption{The ratio $r_{\gamma}(z_s; z_s)$ between the mean shear at redshift $z_s$ of the siren and the shear at redshift $z_{\rm gal} = 2.9$ of the source galaxies. The redshift of the source galaxies is fixed and the $r(z_s,z_{\rm gal})$ is calculated by ${\left\langle \gamma(z_s) \right\rangle}/ {\left\langle \gamma(z_{\rm gal}) \right\rangle}$ to the shear maps in SLICS.}
  \label{fig:ratio_z}
\end{figure}

As Fig.~\ref{fig:correlation} illustrates, the shear values at two different redshifts along the same line of sight are strongly correlated and the shear at lower redshift is smaller on average, so there should be some non-vanishing systematic error introduced by including galaxies with redshift far from that of the siren. Also, Fig.~\ref{fig:correlation} shows that the joint distribution of $\gamma$ at different redshifts along the same line of sight is highly correlated and well approximated by a linear relation. This motivates the expression for the unbiased estimates of the shear at siren's redshift $z_s$ given the shear of the source galaxies with redshift $z_{\rm gal}$, 
\begin{equation} \label{eq:linear_est}
\gamma(z_s)_{|\gamma(z_{\rm gal})} = r(z_s,z_{\rm gal})\gamma(z_{\rm gal}) 
\end{equation}
where $r(z_s,z_{\rm gal}) = {\left\langle \gamma(z_s) \right\rangle}/ {\left\langle \gamma(z_{\rm gal}) \right\rangle}$ is a factor that depends on both the redshift of the siren and the source galaxies. Then the shape noise of the galaxies at different redshifts is scaled by the ratio $r(z_s,z_{\rm gal})$. Fig.~\ref{fig:ratio_z} shows the redshift dependence of the ratio factor $r(z_s,z_{\rm gal})$ with a fixed redshift of the source galaxies at $z_{\rm gal} = 2.9$. 

Galaxies from different redshift bins could provide separate estimations on the shear at the siren's position with errors that are also different from each other. Prior to smoothing and finite truncation of the observation field, the errors in estimating the shear at the siren's position are composed of two parts: the equivalent shape noise $\sigma_{\mu_{s}}$ and the scatter around the linear relation $\sigma_{\mu_{z}}$ given by Eq.~\eqref{eq:linear_est}. The equivalent shape noise $\sigma_{\mu_{s}}$ is given by,
\begin{equation}\label{eq:shape_noise}
\sigma_{\mu_{s}} = r(z_s,z_{\rm gal})\frac{\sigma_{\gamma}}{\sqrt{n_{\rm gal}(z_{\rm gal})}}
\end{equation}
where $\sigma_{\gamma}$ is the effective shape noise for one single galaxy and $n_{\rm gal}$ is the average number of galaxies within one pixel\footnote{Although the galaxy number varies among pixels at the same redshift, the difference is marginal for the purpose of this paper, especially after smoothing.}. According to Eq.~\eqref{eq:linear_est}, $\sigma_{\mu_{s}}$ is sensitive to both the galaxy density $n_{\rm gal}(z)$ and ratio factor $r(z_s,z_{\rm gal})$, leading to different values for different redshift intervals. The scatter around the linear relation $\sigma_{\mu_{z}}$ also has a distinct value for different redshifts, and it has spatial correlation prior to smoothing originated from the intrinsic spatial correlation of the shear at a single redshift.

Because estimations from different redshift bins have different errors, proper weighting must be applied to optimize the integrated results. In principle, the optimal weighting should depend both on $\sigma_{\mu_{s}}$ and $\sigma_{\mu_{z}}$, but the latter error is affected uniquely by smoothing at different redshifts since the intrinsic spatial correlation varies for different redshifts. Fortunately, it turns out that the latter is much smaller than the equivalent shape noise for the smoothing scales and galaxy number densities considered in this paper. Thus we approximate the redshift weighting by,
\begin{equation}\label{eq:redshift_weighting}
w_{z,\gamma}(z_s,z_{\rm gal}) = \frac{1}{\sigma^2_{\mu_{s}}(z_s,z_{\rm gal})}
\end{equation}
where $\sigma_{\mu_{s}}(z_s,z_{\rm gal})$ is given by Eq.~\eqref{eq:shape_noise}. Since the shape noise does not have spatial correlation prior to smoothing and the magnitude of the shape noise is the same for pixels at the same redshift, the smoothing only adds a fixed proportionality constant in front of the equivalent shape noise $\sigma_{\mu_{s}}(z_s,z_{\rm gal})$ at every redshift $z_{\rm gal}$, leading to unchanged weighting after normalization. Therefore, the optimal weighting is independent of smoothing if $\sigma_{\mu_{z}}$ is ignored, which is a valid assumption and simplifies the problem significantly. We will follow this choice of the weighting $w_{z,\gamma}(z_s,z_{\rm gal})$ and ratio factor $r(z_s,z_{\rm gal})$ for the subsequent analyses.

In practice, the weightings should also reflect the image quality of individual galaxies. However, the SLICS catalogs do not emulate the observational details of weak-lensing measurements. Therefore, we neglect technical observation caveats such as the neighbour-exclusion bias~\citep{harnois-derapsCosmologicalSimulationsCombinedProbe2018} as well as the magnification and size bias~\citep{PhysRevD.89.023515}, leaving them to future studies.

\vspace{-1em}
\subsection{Wide-field delensing}\label{wide_delensing}
As introduced in Sec.~\ref{SLICS_intro}, the LSST-like galaxy catalogs can serve well as a test of the delensing procedure using wide-field surveys in the near future. The conditional distribution $p(\mu_{\rm res}|\kappa_{\rm est})$ used to estimate the siren's magnification can be sampled by the simulated galaxies in the LSST-like galaxy catalogs because it is assumed that each siren has a host galaxy (Sec.~\ref{GW_sim:sec}). In principle, galaxies inside the catalogs should be weighted by factors during the construction of $p(\mu_{\rm res}|\kappa_{\rm est})$ depending on the properties of the targeted siren since the properties of the siren are expected to be highly correlated to the properties of its host galaxy (Sec.~\ref{GW_sim:sec} and Sec.~\ref{comb}). These weightings can in principle be used to study the selection effects on sirens with different properties. However, here we only adopt an equal weighting for all galaxies and sirens within the same redshift bin, leaving the discussion on selection effects until the end of Sec.~\ref{Discussion} and middle of Sec.~\ref{Conclusion}.

\begin{figure}
  \graphicspath{ {./ZF_fig/} }
  \includegraphics[width=\linewidth]{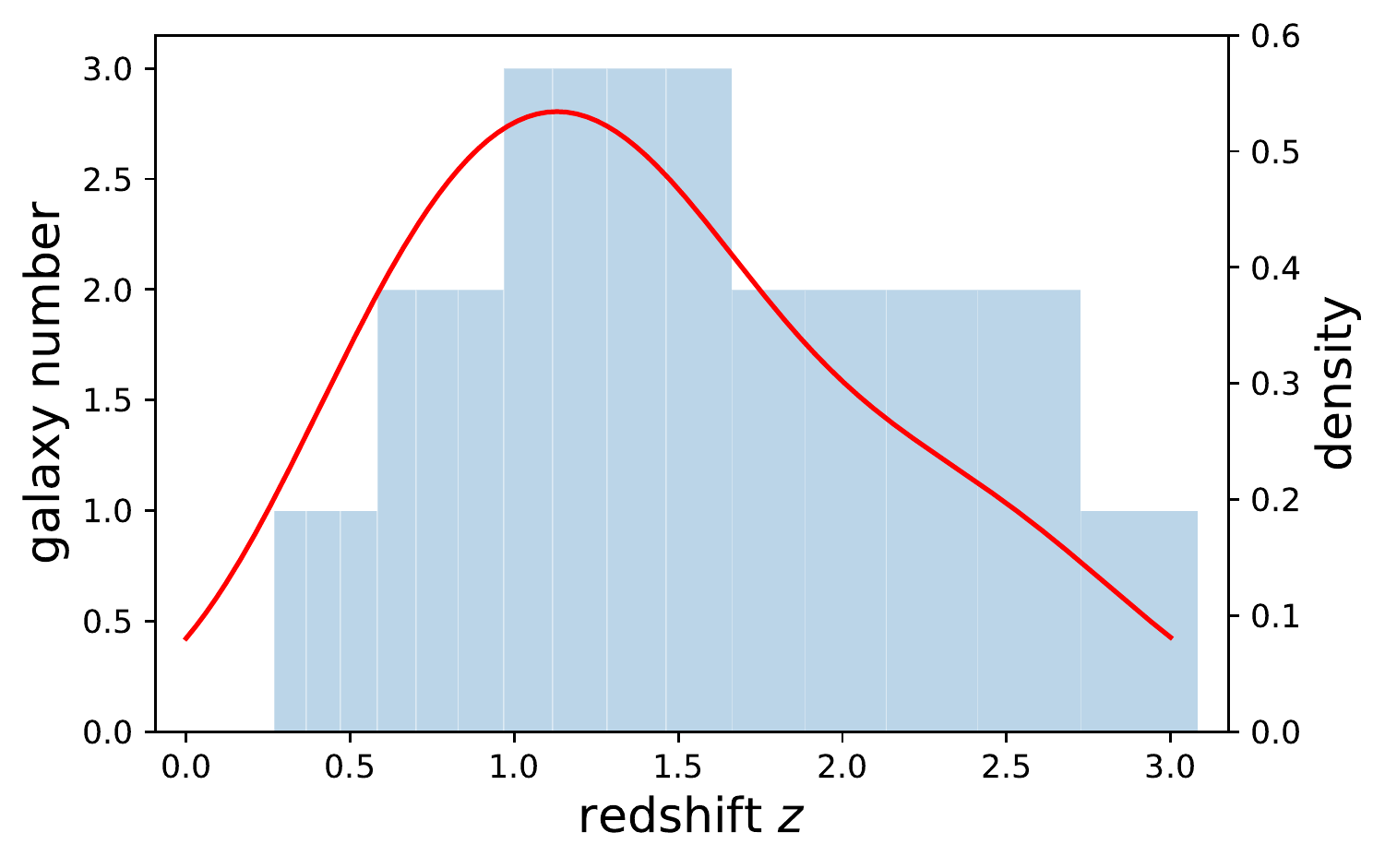}
  \caption{The galaxy distribution $n_{\rm gal}$ along redshift $z$ for the LSST-like catalogs. The galaxy number denotes the total number of galaxies within one square arcminute and redshift bin. The red curve shows the corresponding density distribution function. The peak of the distribution is between $z = 1$ and $z = 1.5$.}
  \label{fig:wide_gal_distribution}
\end{figure}

The key features related to the reconstruction behaviours are listed in Table~\ref{tab:LSST}. Additionally, the redshift dependence of the galaxy number per ${\rm arcmin}^2$ is shown in Fig.~\ref{fig:wide_gal_distribution}. To emulate what happens in practice, the galaxies in the LSST-like catalogs are divided into 17 redshift bins according to their true redshift. The choice of the bin number and bin edges are consistent with the SLICS lensing simulation setup \citep{harnois-derapsCosmologicalSimulationsCombinedProbe2018}, where the bin edges are placed at the redshifts of the source planes $z_{\rm source}$ in Table~\ref{tab:z_planes}.

In more realistic situations, only the photometric redshifts of the galaxies are accessible and thus it would be possible to have the galaxies categorized into the wrong redshift bins. Here we ignore this effect and do not explore the optimal choice of the redshift bins. The exploration of the optimal choice of redshift bins should rely on some weak-lensing simulations that do not have the same set of special redshifts for all realizations (such as different sets of $z_{\rm source}$ and $z_{\rm lens}$) and treat the weak-lensing properties along redshift more seriously, i.e. not just linear interpolation.

\begin{figure}
  \graphicspath{ {./ZF_fig/} }
  \includegraphics[width=\linewidth]{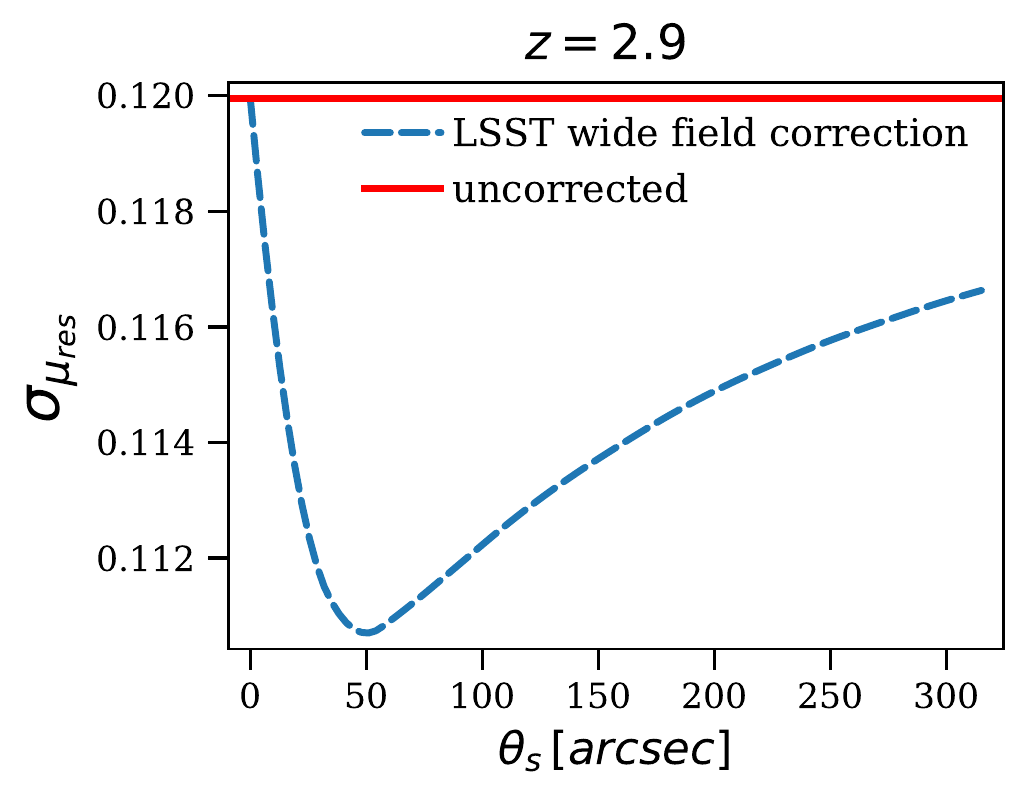}
  \caption{The dispersion of residual magnification $\sigma_{\mu_{\rm res}}$ (dashed line) for sirens at redshift $z_s = 2.9$ as a function of the smoothing filter scale $\theta_s$ for an LSST-like wide-field survey. The solid horizontal line marks the magnification dispersion without correction. The optimal smoothing filter scale is around $50$ arcsecond with a reduction of the residual dispersion being less than 10\%.}
  \label{fig:wide_smoothing}
\end{figure}

\begin{figure}
  \graphicspath{ {./ZF_fig/} }
  \includegraphics[width=\linewidth]{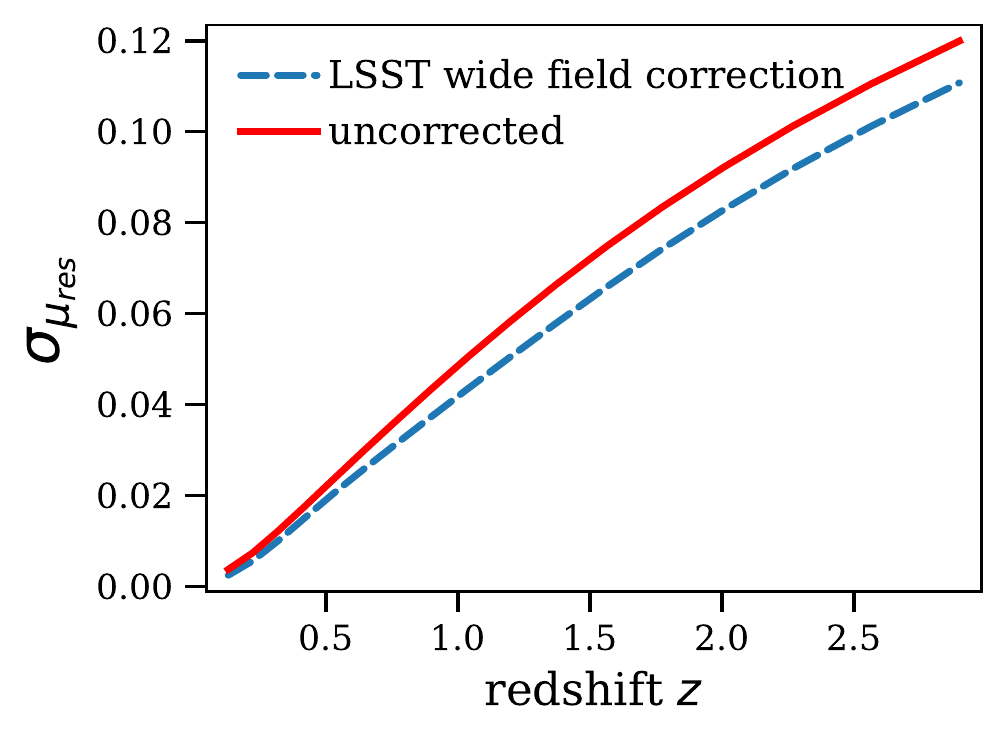}
  \caption{The dispersion of residual magnification $\sigma_{\mu_{\rm res}}$ (dashed line) as a function of the siren's redshift $z$. The dispersion without
  correction ($\mu_{\rm est} = 1$, solid line) is compared to the residual dispersion with convergence reconstructed from an LSST-like wide-field survey under the optimal filter scale (dashed line). The weak-lensing errors are only reduced by about 10\% at all redshifts.}
  \label{fig:wide_redshift}
\end{figure}

In another more realistic scenario, the observed ellipticities of the galaxies in the mock catalogs are noisy due to measurement error and shape noise. Thus, certain smoothing must be applied to reduce the noise level. Fig.~\ref{fig:wide_smoothing} quantifies how much the dispersion of the residual magnification $\sigma_{\mu_{\rm res}}$ for a standard siren at $z_s=2.9$ can be reduced with an LSST-like futuristic wide-field survey for particular choices of the filter scale $\theta_s$. With the optimal filter scale $\theta_s \approx 50$ arcsec, Fig.~\ref{fig:wide_redshift} then shows how $\sigma_{\mu_{\rm res}}$ changes as a function of the siren's redshift $z_s$. As Fig.~\ref{fig:wide_redshift} illustrates, the weak-lensing errors are only reduced by about 10\% compared to the uncorrected case at all redshifts. There is little benefit from delensing using an LSST-like wide-field survey. 

Since the total area of the field is large enough ($10 \times 10\,{\rm deg}^2$), the mass-sheet degeneracy problem is negligible. The poor performance of the LSST-like wide-field surveys mainly originates from the inadequate number density of the galaxies. Since the equivalent shape noise $\sigma_{\mu_{s}}$ within one pixel depends on the galaxy density and can only be suppressed by smoothing, an insufficiently high galaxy density will lead to a large filter scale $\theta_s$ to obtain an acceptable signal-to-noise ratio for our magnification estimate. The large filter scale smooths away the small-scale contributions to the magnification, resulting in very poor performance of delensing. To significantly improve the delensing result, we therefore must have a deep and dedicated survey that points to the targeted siren.

\subsection{Golden siren selection}\label{golden}

The end of Sec.~\ref{wide_delensing} demonstrates that an LSST-like futuristic wide-field survey is not sufficient to reduce the weak-lensing error significantly. Therefore, we must turn to a deep-field survey with sufficient resolution and depth to obtain images with adequate galaxy number density. Because such a deep-field survey is very expensive in practice, it may only be feasible to conduct the deep-field survey around one specific siren.

We refer to the best siren inside a catalog for carrying out the delensing process as a `golden siren'\footnote{This terminology has already been introduced in the literature \citep{2013ApJ...767..124N} to refer to a particularly well-localized, nearby bright siren, and we adapt the term to the current context.}. The siren is chosen based on how well the delensing process can improve the accuracy of parameter estimation, after reducing the dispersion of the residual magnification for the siren. The performance of the siren is affected by the measurement error on its estimated luminosity distance and the redshift of the siren. With a smaller measurement error, the siren can better constrain the cosmological parameters. On the other hand, given the same measurement errors, a siren can generally better constrain the cosmological parameters when the siren is further away, according to Eq.~\eqref{lum_z}. Therefore, we only consider sirens with $z_s > 1$ as candidates for being the `golden siren'. We note, however, that the siren with the highest redshift may not have the smallest measurement errors. The trade-off between smaller measurement errors and higher redshift is complicated. We attempt to reweight the parameter estimation results using the likelihood of individual sirens with reduced $\sigma_{\mu_{\rm res}}$ to select the golden siren from each of our siren catalogs, and investigate the improvements in cosmological parameter estimation for different $\sigma_{\mu_{\rm res}}$.

The likelihood of parameter estimation after correction can be expanded as
\begin{equation}
\begin{split}
    & p(d_1\cdots d_i^{\text{reduced $\sigma_{\mu_{\rm res}}$}}\cdots d_n|\Vec{\Omega}) \\
    &\propto p(d_1\cdots d_n|\Vec{\Omega})\times \frac{p(d_i^{\text{reduced $\sigma_{\mu_{\rm res}}$}}|\Vec{\Omega})}{p(d_i|\Vec{\Omega})}\,,
\end{split}
\end{equation}
where $p(d_i^{\text{reduced $\sigma_{\mu_{\rm res}}$}}|\Vec{\Omega})/p(d_i|\Vec{\Omega})$ is the factor used to reweight the estimation results. Here $d_i$ is the GW data of the chosen siren, and $d_{j\neq i}$ is the GW data of other sirens. $\Vec{\Omega}$ is the vector of cosmological parameters and other information such as $\mathbf{z}_{\rm s}$ is not shown explicitly.

By comparing the contour area in the posterior of the cosmological parameters, one can evaluate quantitatively how the reduction of $\sigma_{\mu_{\rm res}}$ for the chosen siren improves the cosmological parameter estimation.

\begin{figure}
    \centering
    \includegraphics[width=\linewidth]{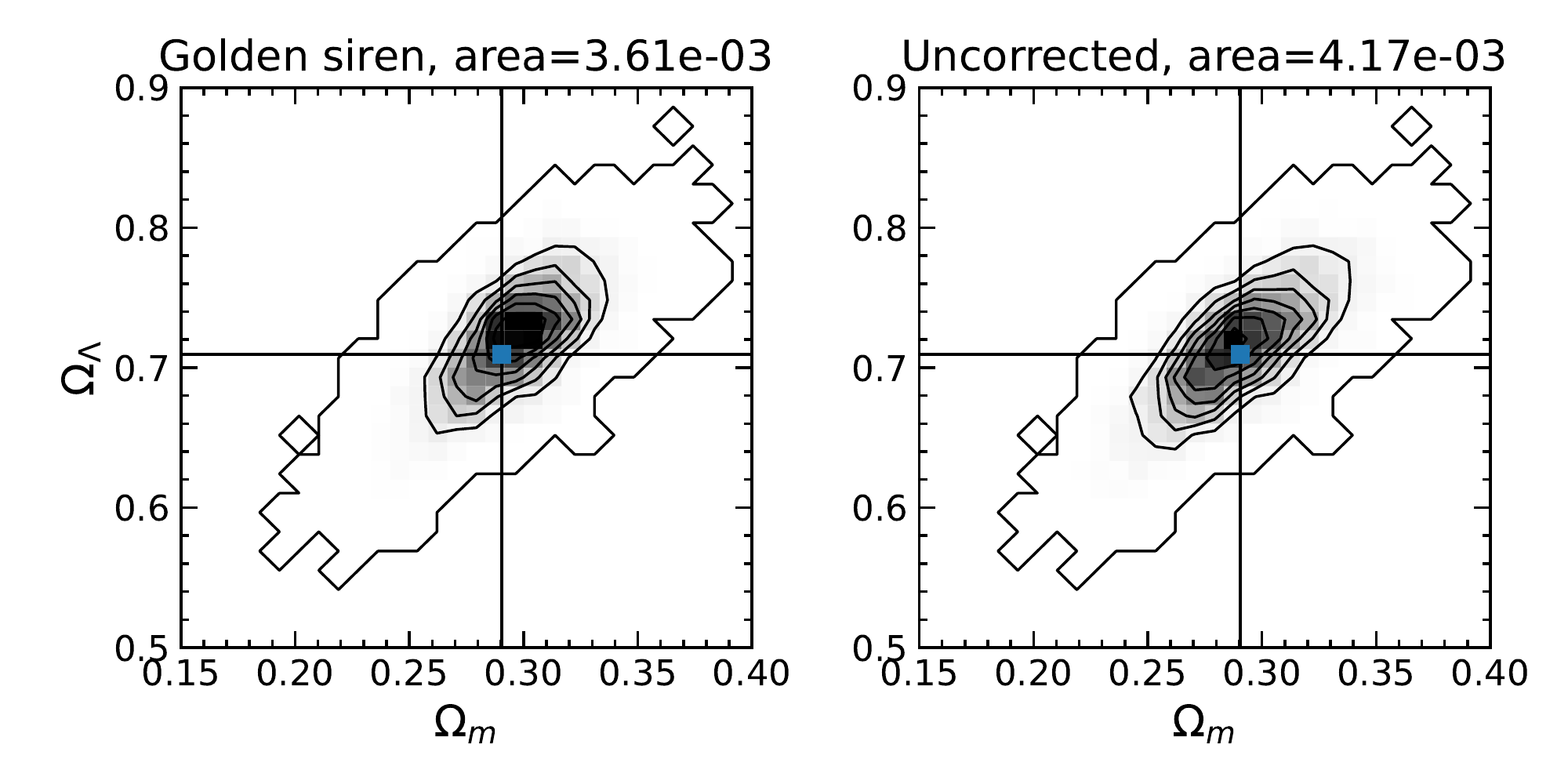}
    \includegraphics[width=\linewidth]{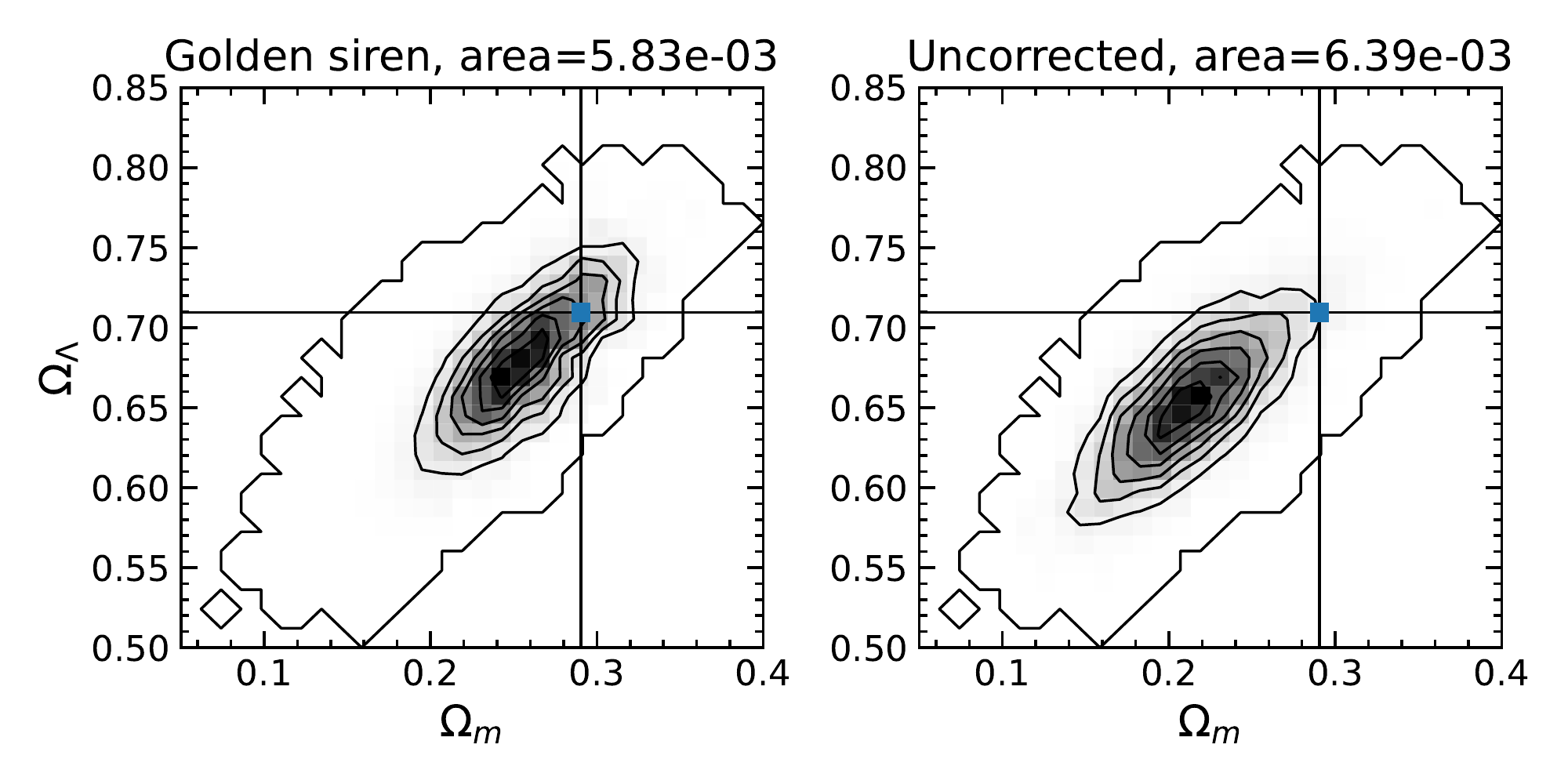}
    \includegraphics[width=\linewidth]{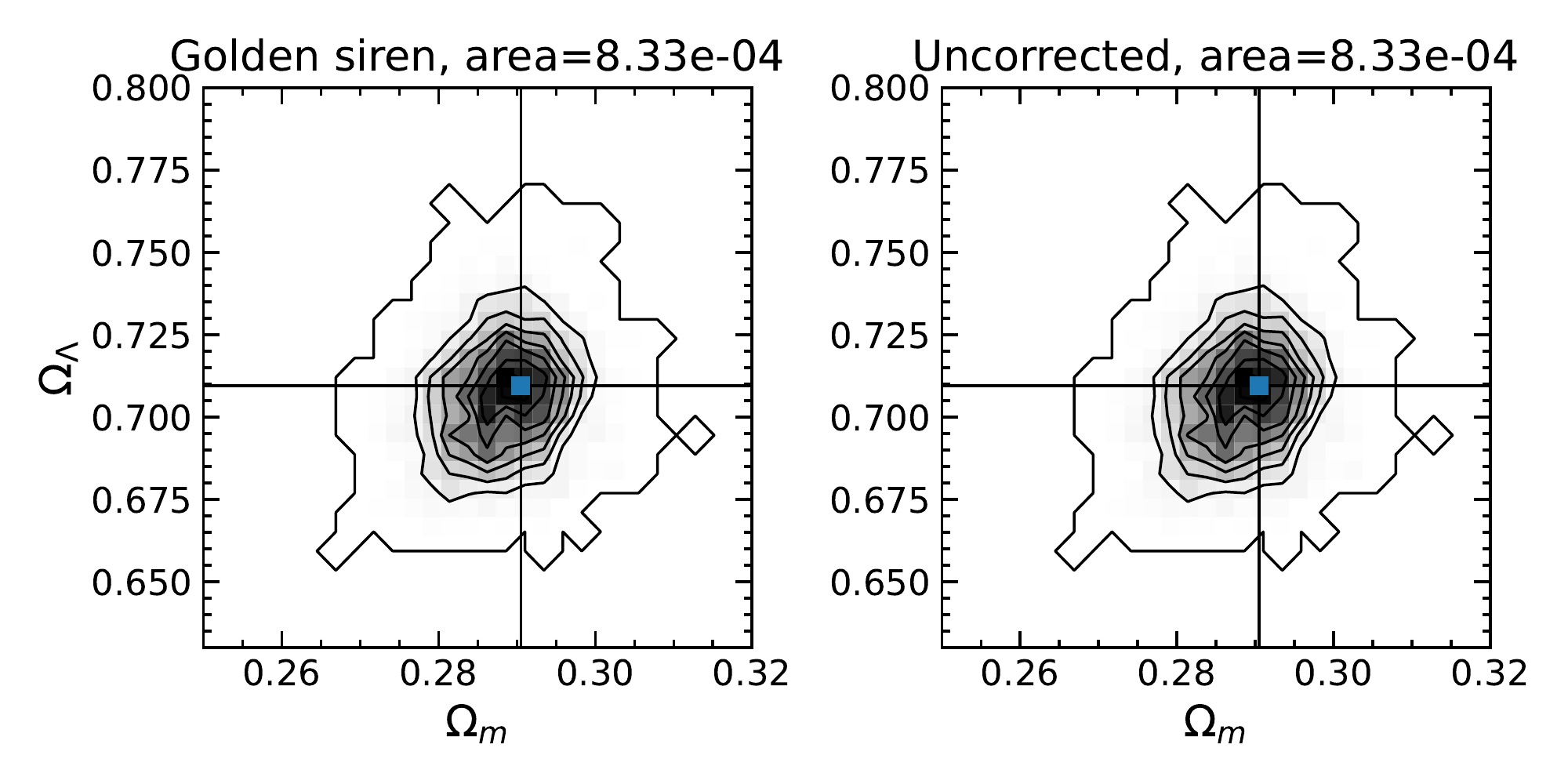}
    \caption{Examples of credible region contours for the inferred cosmological parameters before (labelled `Uncorrected', right panels) and after (labelled `Golden siren', left panels) delensing of the golden siren identified in each mock siren catalog. Examples are shown for each of the three formation models considered: popIII (top), Q3d (mid), Q3nod (bottom). The residual dispersion is fixed to $\sigma_{\mu_{\rm res}} = 5\%$, i.e. a factor of two lower than the uncorrected $\sigma_{\mu} \approx 10\%$.
    The blue square shows the true values of the cosmological parameters. The change in the contour areas of the credible regions is marginal and insignificant.
    }
    \label{fig:compare_contour}
\end{figure}

\begin{figure*}
    \centering
    \includegraphics[width=.32\linewidth]{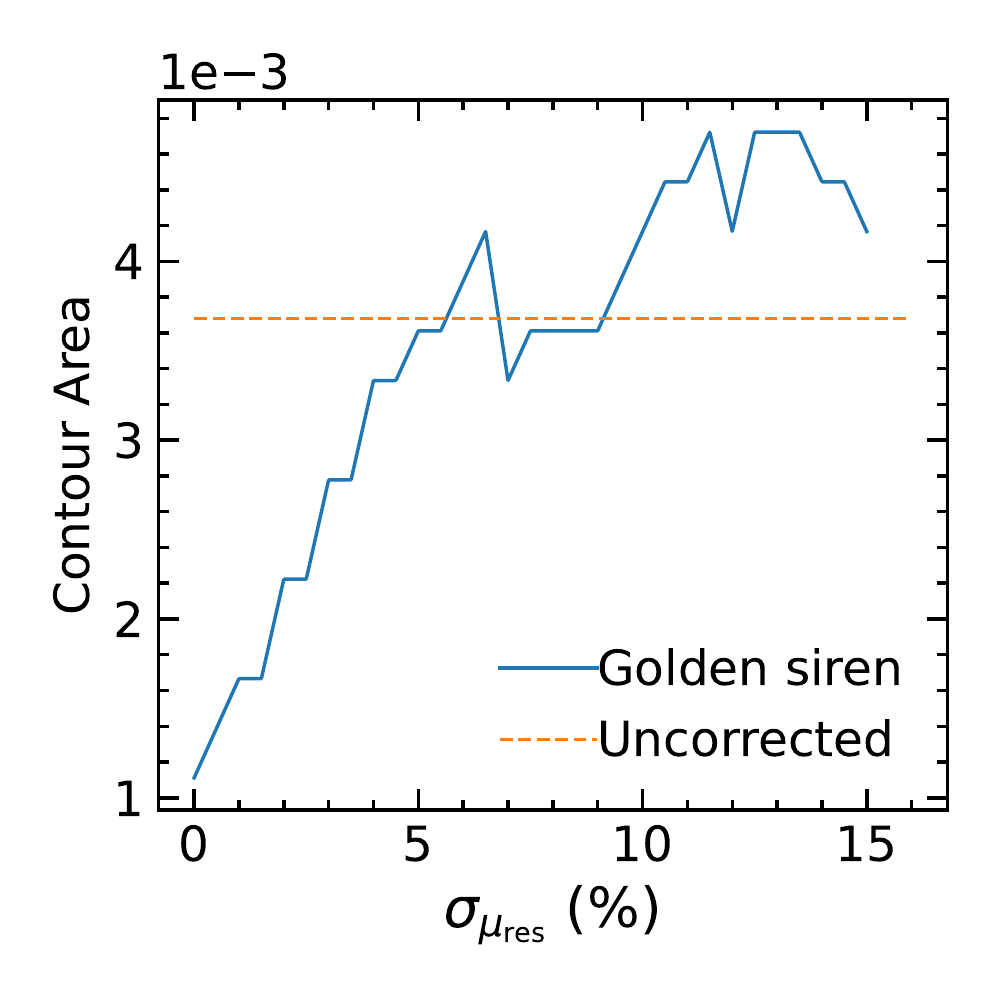}\includegraphics[width=.32\linewidth]{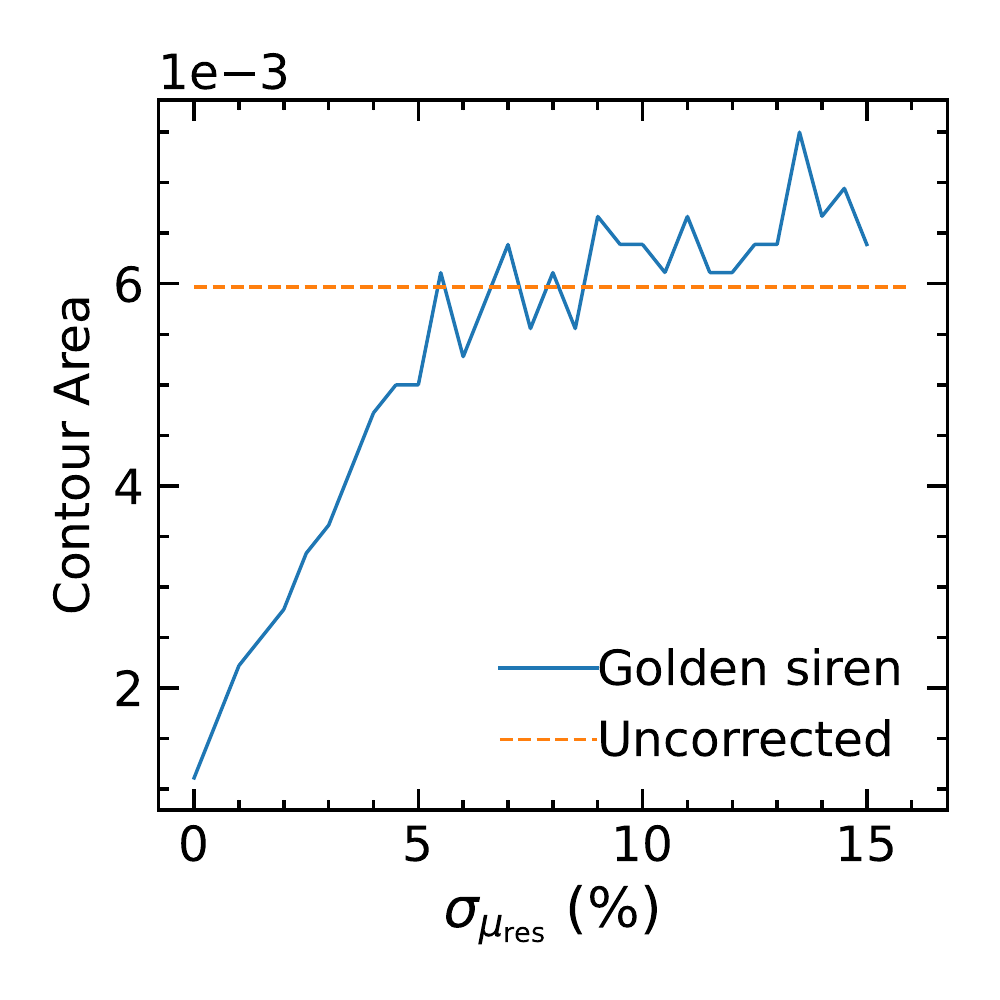}\includegraphics[width=.32\linewidth]{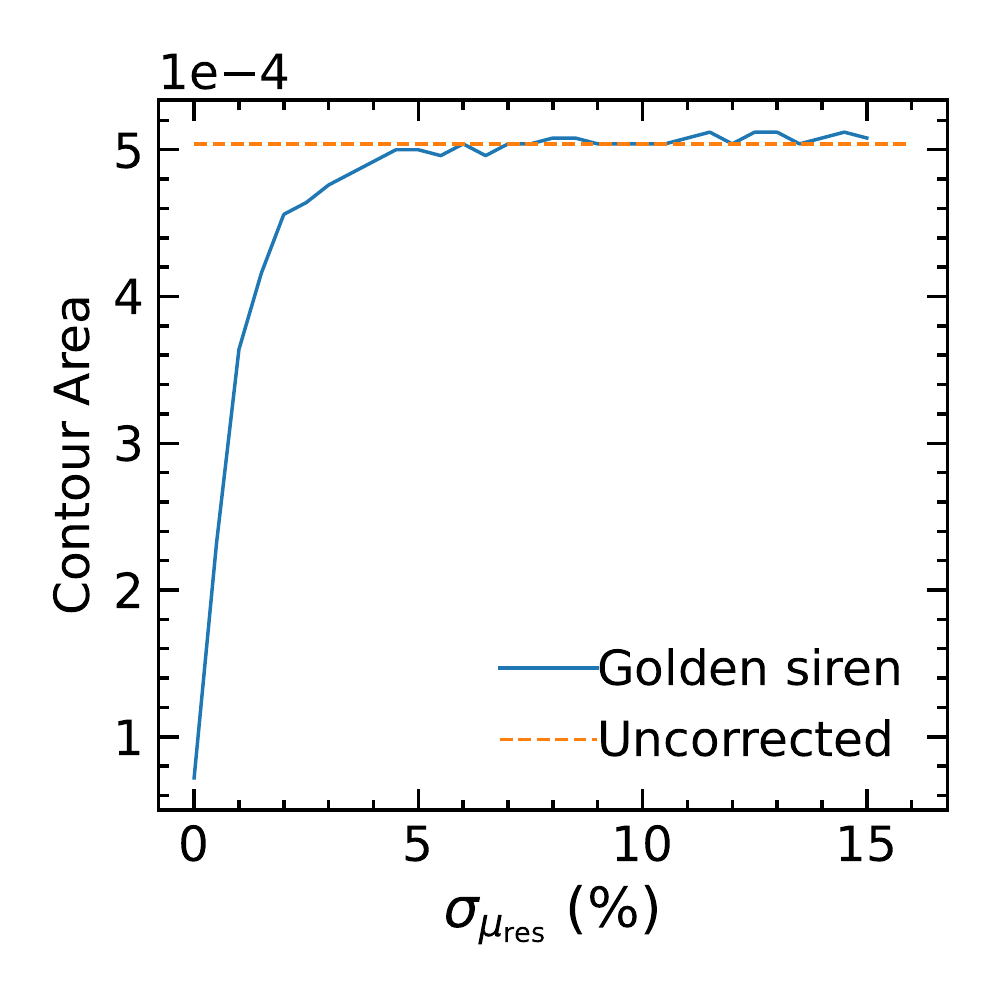}
    \caption{The area of the credible region contour as a function of the dispersion of residual magnification $\sigma_{\mu_{\rm res}}$ for the delensed golden siren under formation models popIII (left), Q3d (mid) and Q3nod (right). The horizontal dotted line denotes the area in the uncorrected case. The plots show the increasing trend of the contour area with growing $\sigma_{\mu_{\rm res}}$. Random Poissonian errors from the sampling processes are responsible for the fluctuations at high $\sigma_{\mu_{\rm res}}$. To improve the cosmological parameter estimation significantly (factor of two or greater reduction), the residual dispersion $\sigma_{\mu_{\rm res}}$ for the golden siren after delensing should be less than 2\%. 
    }
    \label{fig:area_vs_error}
\end{figure*}

Fig.~\ref{fig:compare_contour} shows examples of the credible regions obtained for the cosmological parameters\footnote{We show here contours for the dimensionless density parameters only, in the idealised case where the value of the Hubble constant is considered to be known already.} before (labelled `Uncorrected', shown in the right panels) and after (labelled `Golden siren', shown in the left panels) delensing of the golden siren identified in each mock siren catalog. Example results are shown for each of the three formation models considered in our analyses: popIII (top), Q3d (mid), Q3nod (bottom). In each case the residual dispersion is fixed to $\sigma_{\mu_{\rm res}} = 5\%$, i.e. a factor of two lower than the uncorrected $\sigma_{\mu} \approx 10\%$. These plots show that reducing $\sigma_{\mu_{\rm res}}$ for the golden siren can not reduce the contour area of the posterior of the cosmological parameters appreciably, as the change is marginal and insignificant. Moreover we note, again, that these results are for the case where the Hubble constant is fixed to equal its true value -- i.e. this parameter is considered to be already known (or very tightly constrained) from other observations. Even in this idealised case, the delensing of the golden siren does not result in any significant reduction in the area of the credible region for the other cosmological parameters. In practice, the uncertainty in $H_0$ can be considered as an extra error on the estimated luminosity distance and the effect of delensing is more insignificant.

Fig.~\ref{fig:area_vs_error} shows how the area of the credible region contour inferred for the cosmological parameters changes with increasing dispersion of the residual weak-lensing magnification. The left, middle and right plots are for the popIII, Q3d and Q3nod models respectively. The overall increasing trend of the contour area as the dispersion increases is clear, although there are some random Poissonian errors from the sampling processes for high residual dispersions. However, even after delensing the golden siren, the contour area is comparable in size to the uncorrected case when $\sigma_{\mu_{\rm res}} \approx 5\%$, and shows a significant (factor of two or greater) reduction in size only if $\sigma_{\mu_{\rm res}} \le 2\%$. The feasibility of reducing $\sigma_{\mu_{\rm res}}$ to such a level using convergence maps reconstructed from deep-field surveys is evaluated quantitatively in the following sections.

\subsection{Deep-field delensing}
\subsubsection{Construction of deep-field survey}
The SLICS simulations do not contain deep-field mock galaxy catalogs, so the construction of such catalogs proved necessary. To emulate state-of-the-art technologies, here we adopt a configuration similar to the \textit{James Webb Space Telescope} (JWST) ultra-deep field (UDF) for the construction of the deep-field survey. 

\begin{figure}
  \graphicspath{ {./ZF_fig/} }
  \includegraphics[width=\linewidth]{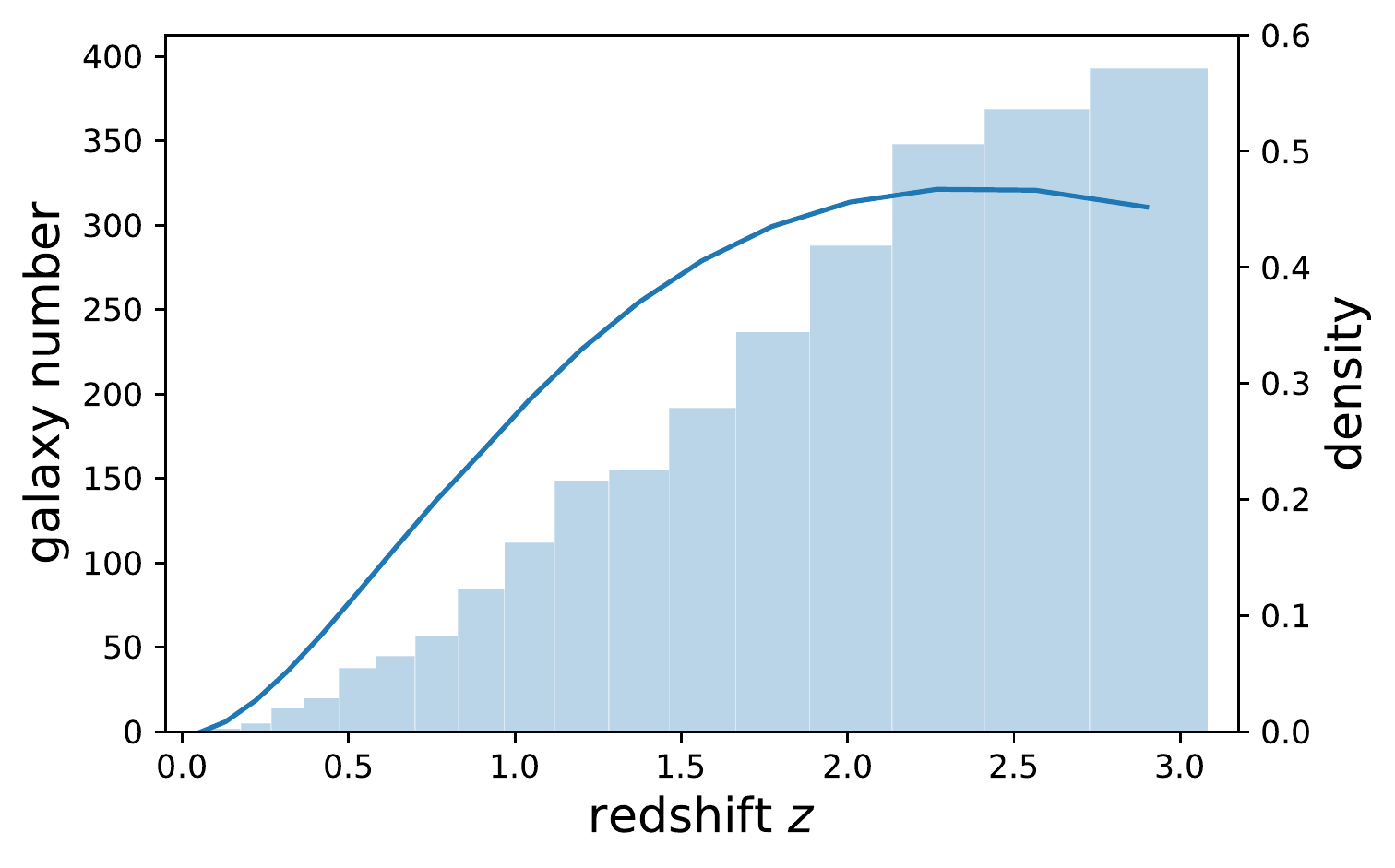}
  \caption{The galaxy distribution $n_{\rm gal}$ as a function of redshift $z$ for the JWST-like galaxy catalogs used in our study. The galaxy number denotes the total number of galaxies within one square arcminute and redshift bin. The blue curve shows the corresponding density distribution function. There is a sudden truncation at $z=3$, aligning with the maximal redshift for lensing maps in SLICS.}
  \label{fig:deep_gal_distribution}
\end{figure}

The observational setup and the redshift distribution of galaxies that we adopt follow the simulated JWST UDF catalogs \citep{yungSemianalyticForecastsJWST2022a}. Since the satellites might be too dim to perform accurate measurements of their shapes, only the field galaxies within the catalogs are counted towards the galaxy number density. Fig.~\ref{fig:deep_gal_distribution} shows how the galaxy density (i.e. galaxy number within one ${\rm arcmin}^2$ and redshift bin) changes with redshift in our simulated deep-field survey. Note that the measurement errors in galaxy shape for the JWST UDF are complicated and small compared to the shape noise. For simplicity, therefore, only the shape noise will be counted in determining the shear dispersion per galaxy, i.e. $\sigma_\gamma = \sigma_{\rm shape} = 0.26$ \citep{schaanLookingSameLens2017}. Other information about the simulated deep-field survey is summarized in Table~\ref{tab:JWST}. Note that the underlying weak-lensing maps are the same as in the LSST-like wide-field survey.

To make a consistent construction of the deep-field survey with respect to the lensing maps, we produced deep-field mock galaxy catalogs so that the positions of the galaxies trace the underlying dark matter distribution with a fixed bias. We do this by utilizing the projected 2D density contrast $\delta_{2D}(\boldsymbol{\theta},z)$ defined by Eq.~\eqref{eq:2D_density_contrast} such that the density distribution of galaxies in each redshift bin is proportional to the mass distribution projected within the bin \citep{harnois-derapsCosmologicalSimulationsCombinedProbe2018}. The proportionality constant is the linear galaxy bias which is assumed to be $b = 1.18$, the same as the bias in the KiDS-HOD mocks included in SLICS.\footnote{Both the LSST-like mock catalogs and the KiDS-HOD mocks are straightforward extensions of a single HOD model.} The redshifts of the galaxies within a bin follow a random uniform distribution for each line-of-sight.

This construction has the advantage that both the lensing maps and deep-field galaxy catalog arise from the same set of density contrasts $\delta_{2D}(\boldsymbol{\theta},z)$. However, the bias here is a fixed parameter instead of being redshift, scale and mass dependent. The linear bias model might also not be sufficient for explaining the small-scale galaxy/matter distribution. Future studies are necessary to investigate these effects.

The augmented galaxies added to our deep-field catalog could be interpreted as galaxies too faint to be observed in a wide-field survey. However, since the approach to simulate the augmented galaxies is too simplistic, these galaxies are not considered as the potential host galaxies for SMBHBs and hence not viewed as samples towards the distributions $p(\mu_{\rm res}|\kappa_{\rm est})$. This issue will be further discussed in Sec.~\ref{Conclusion}. 

\begin{table*}
    \centering
    \begin{tabular}{c|c|c|c|c}
    \hline
    Total area of the field & The redshift range &	Galaxy number density & Resolution of lensing maps & Shear dispersion (single galaxy) \\
    \hline
    $11 \times 11 \; {\rm arcmin}^2$ & $0\,-\,3$ & $ \approx \,2500\,{\rm arcmin}^{-2}$ & $\approx\;4.6\:{\rm arcsec}$ & $\sigma_\gamma = 0.26$\\
    \hline
    
    \end{tabular}
    \caption{Relevant information about the JWST-like deep-field catalog used in this study.}
    \label{tab:JWST}
\end{table*}

\vspace{-1em}
\subsubsection{Deep-field delensing results}
The high number density of galaxies strongly suppresses the shape noise and only a small smoothing scale is required to obtain an acceptable signal-to-noise ratio. This is very beneficial for extracting small-scale signals in reconstruction and thus allows one to substantially reduce the residual dispersion $\sigma_{\mu_{\rm res}}$ for standard sirens. In Sec.~\ref{golden}, the golden siren is assumed to have redshift $z_s > 1$ and hence the distribution $p(\mu_{\rm res}|\kappa_{\rm est})$ will be investigated at $z>1$ only.

\begin{figure}
  \graphicspath{ {./ZF_fig/} }
    \includegraphics[width=\linewidth]{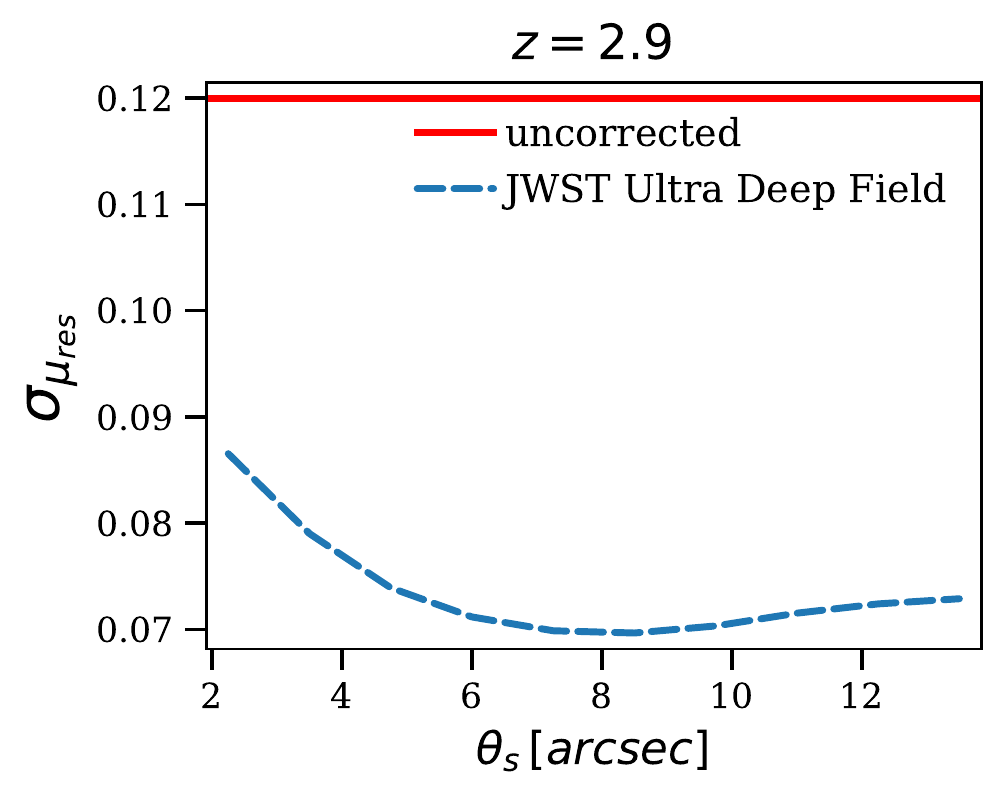}
  \caption{The dispersion of residual magnification $\sigma_{\mu_{\rm res}}$ (dashed line) for sirens at redshift $z_s = 2.9$ as a function of the smoothing filter scale $\theta_s$ from a JWST-like deep-field survey. The solid horizontal line marks the magnification dispersion without correction. The optimal smoothing filter scale is around $8$ arcsecond, resulting in a 42\% reduction of the residual dispersion.}
  \label{fig:deep_smoothing}
  \includegraphics[width=\linewidth]{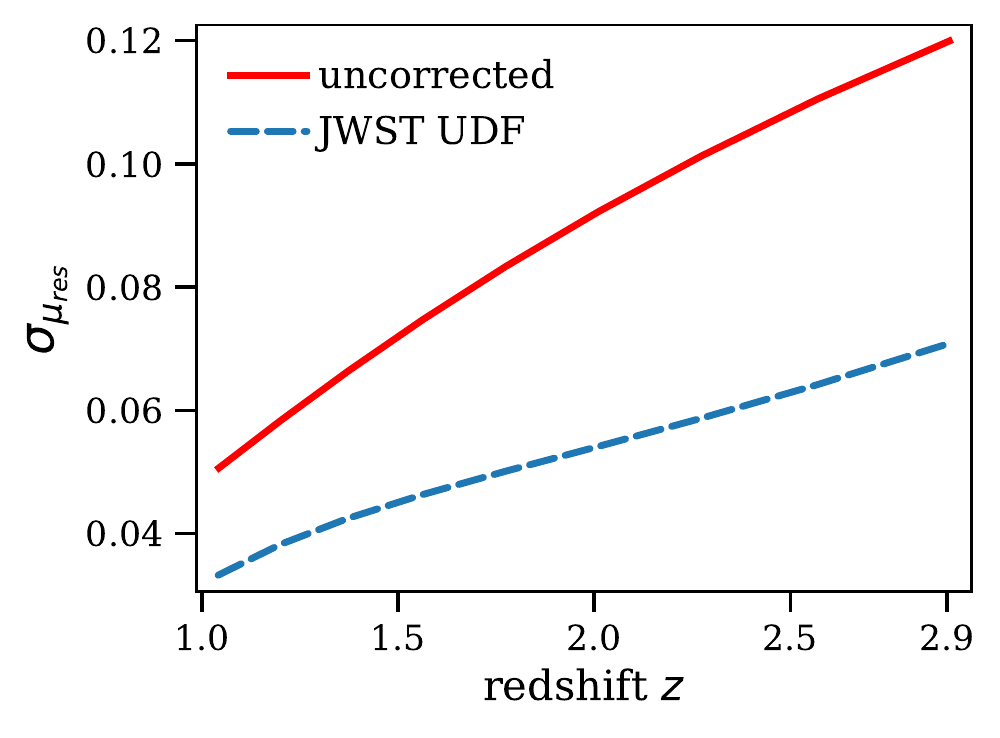}
  \caption{The dispersion of residual magnification $\sigma_{\mu_{\rm res}}$ (dashed line) as a function of the siren's redshift $z$. The dispersion without
  correction ($\mu_{\rm est} = 1$, solid line) is compared to the residual dispersion with convergence reconstructed from a JWST-like deep-field survey under the optimal filter scale (dashed line). The weak-lensing errors are reduced by around $30 \sim 40 \%$ at all redshifts, significantly better than the outcomes from LSST-like wide-field surveys.}
  \label{fig:deep_redshift}
\end{figure}

As Fig.~\ref{fig:deep_smoothing} shows, the residual dispersion $\sigma_{\mu_{\rm res}}$ can be reduced by delensing to around $60\%$ of the uncorrected dispersion at $z_s = 2.9$ with the optimal filter scale $\theta_s \approx 8$ arcsec. Under the best filter scale $\theta_s \approx 8$ arcsec, the redshift dependence of the residual dispersion $\sigma_{\mu_{\rm res}}$ is shown in Fig.~\ref{fig:deep_redshift}. Obviously, the delensing outcomes are significantly better than the LSST-like wide-field surveys.

\begin{figure}
  \graphicspath{ {./ZF_fig/} }
  \includegraphics[width=\linewidth]{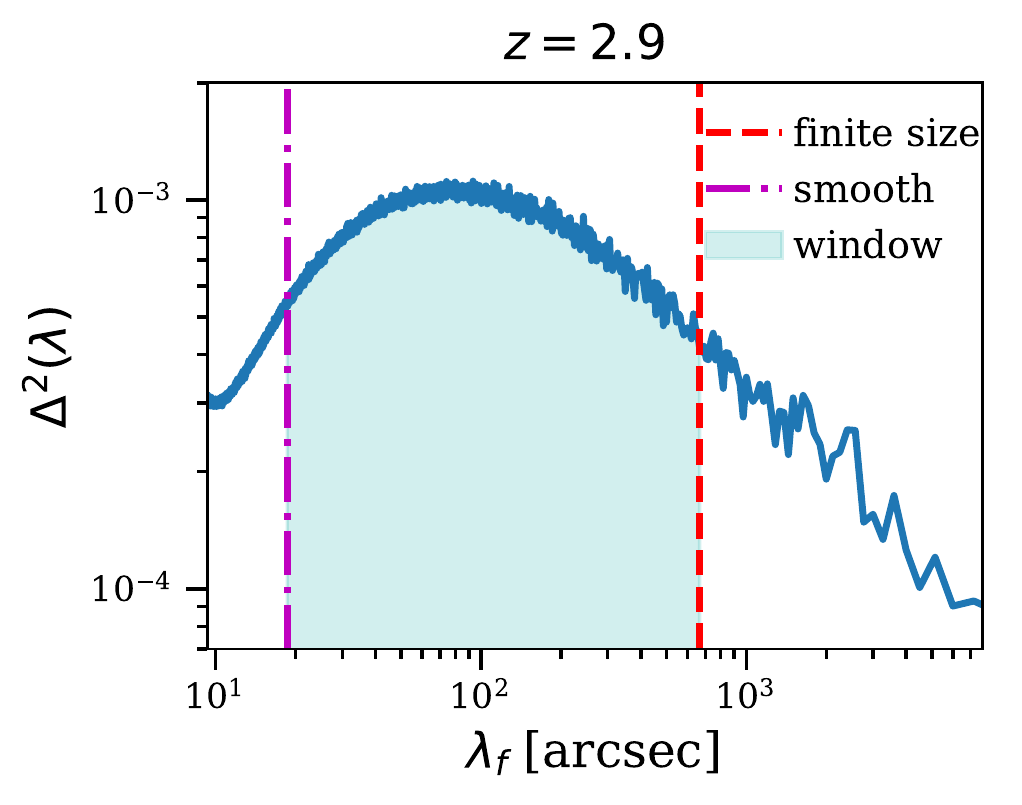}
  \caption{The logarithmic band power of the convergence field at $z = 2.9$. The area under the curve at different intervals of the wavenumbers $\lambda_f$ for the fluctuation modes quantifies their contribution to the total variance. The fluctuation modes with wavenumbers $\lambda_f \approx 80$ arcsec contribute the most. The detection window for the fluctuation modes limited by smoothing (lower limit) and finite size of the field (upper limit) is also denoted. Note that the log scale in the $x$ axis has been compensated so that the proportion of the area under the curve is the same as in linear scale. However, no manipulations have been carried out on the $y$ axis.}
  \label{fig:logarithmic_bandpower}
\end{figure}

Apart from smoothing, which wipes away the small-scale fluctuations, the mass-sheet degeneracy originating from the limited size of the deep survey also contributes to the residual dispersion $\sigma_{\mu_{\rm res}}$. The size of the JWST UDF is much larger than the Hubble Space Telescope UDF, but it is still not sufficient for the mass-sheet degeneracy problem to be negligible. Basically, the smoothing and finite size of the observation field put lower and upper bounds on the detectable scale in convergence fluctuations. Fig.~\ref{fig:logarithmic_bandpower} illustrates the logarithmic band power of the convergence field at $z = 2.9$, which illustrates the contribution at different fluctuation scales to the total convergence variance by the area below the curve. The best window of detection is also denoted in Fig.~\ref{fig:logarithmic_bandpower}. 

Although the peak in the logarithmic band power of convergence has been included in the window, signals out of the detection window are not inconsiderable, especially the part from the large-scale fluctuations. Fortunately, as mentioned in Sec.~\ref{mass_sheet}, fluctuations larger than the deep survey area can be partially eliminated by an accompanying wide-field survey. The large-scale fluctuations beyond the size of the deep images can be picked up by a wide-field survey that has an adequate field size. We will present the method and results of this combination in the following section. 

\vspace{-1em}
\subsection{Hybrid observation delensing}\label{hybrid}
Sec.~\ref{wide_delensing} demonstrated that LSST-like wide-field surveys are blind to sub-arcminute convergence fluctuations that would require significantly high galaxy densities to probe them. However, they have the potential to detect fluctuations larger than a few tens of arcminutes since wide-field surveys of this type have enough width and the shape noise problem is less serious for these scales of consideration. To take full advantage of the potential of delensing, we can therefore estimate the siren's magnification from a hybrid observation composed of both deep- and wide-survey data, similar to the composite approach considered in \citet{shapiroDelensingGravitationalWave2010b}. The wide survey is dedicated to probing fluctuation modes larger than the size of the deep image and the deep survey is able to recover fine features. 

Although the underlying matter fluctuations are non-Gaussian in principle, we assumed that the fluctuation modes larger than the size of the deep survey do not couple to the modes within the deep image. This can be partially justified by the relatively large area of the JWST-like UDF ($11 \times 11 \; arcmin^2$) where the fluctuations beyond the size of the UDF are within an almost linear regime and well approximated by a Gaussian random field. Future studies should investigate this caveat and how it depends on the specific choice of the deep field survey.

In this scenario, we are unconcerned about small-scale fluctuations in the LSST-like wide-field survey. Therefore, the pixel size used in reconstruction for the dedicated wide-field survey should be the same as the field size of the deep survey. Then the number of galaxies within one pixel increases dramatically, making the shape noise less severe. 

However, the galaxy density may not be sufficient to obtain an acceptable signal-to-noise ratio even under this chosen pixelation and certain smoothing can be helpful to reduce the noise level. To quantify the ability of a wide-field survey for removing the mass-sheet degeneracy of the deep survey, we defined the residual mass-sheet degeneracy as,
\begin{equation}
\mu_{\rm res(ms)} = \mu_{\rm ms} - \mu_{\rm est(ms)}
\end{equation}
where $\mu_{\rm ms}$ is the part of magnification contributed from the fluctuation modes larger than the deep survey size and $\mu_{\rm est(ms)}$ is the estimated value for $\mu_{\rm ms}$ from the wide-field survey. 

\begin{figure}
  \graphicspath{ {./ZF_fig/} }
  \includegraphics[width=\linewidth]{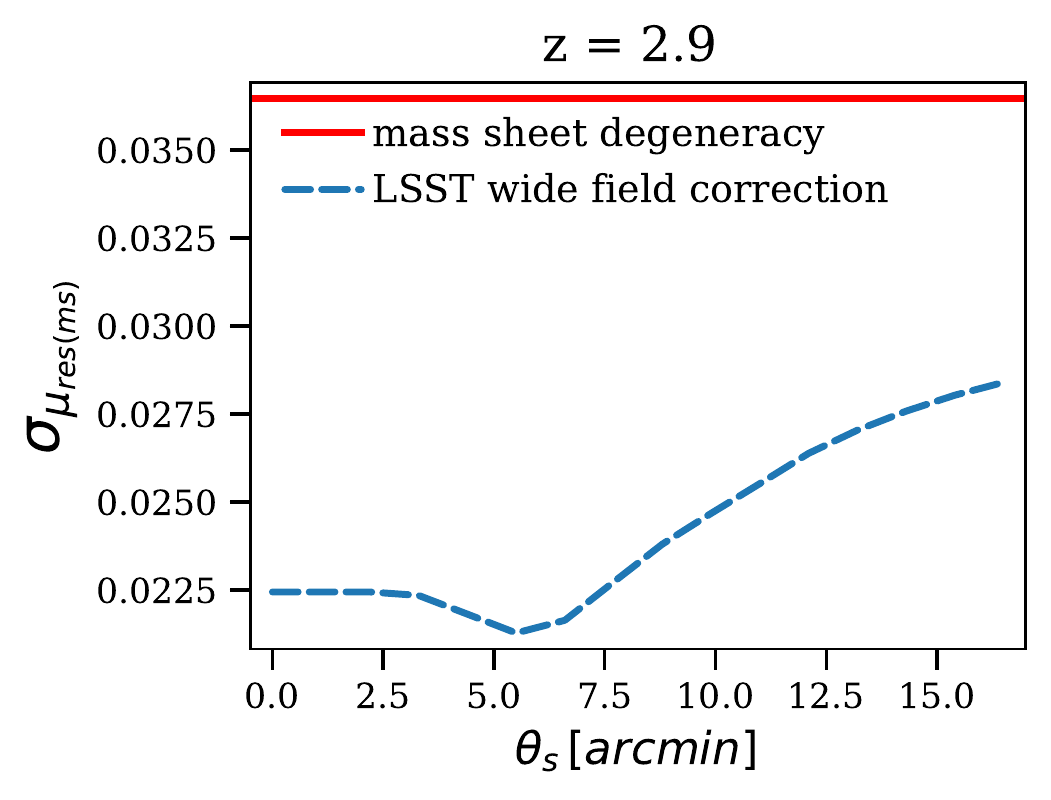}
  \caption{The dispersion of the residual mass-sheet degeneracy $\sigma_{\mu_{\rm res(ms)}}$ (dashed line) for sirens at redshift $z_s = 2.9$ as a function of the smoothing filter scale $\theta_s$ from an LSST-like wide-field survey. The solid horizontal line marks the residual mass-sheet degeneracy without correction. The optimal smoothing filter scale is $\approx 5.5$ arcminute with a $ \approx 40\%$ reduction.}
  \label{fig:sm_mass_sheet_correction}
\end{figure}

Fig.~\ref{fig:sm_mass_sheet_correction} quantifies how the dispersion of the residual mass-sheet degeneracy $\sigma_{\mu_{\rm res(ms)}}$ changes as a function of the filter scale $\theta_s$ for a siren located at $z=2.9$. It illustrates that the mass-sheet degeneracy uncertainty in a JWST-like deep-field survey can be reduced to $\sim 60\%$ with the help of an LSST-like wide-field survey. 

\begin{figure}
  \graphicspath{ {./ZF_fig/} }
  \includegraphics[width=\linewidth]{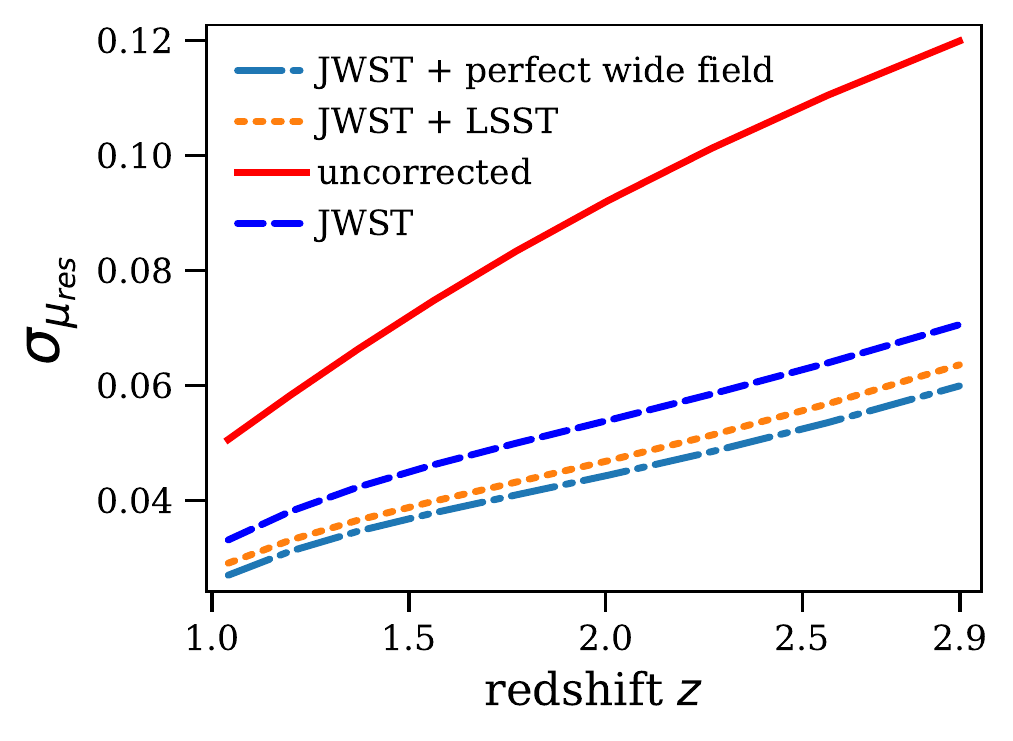}
  \caption{The dispersion of residual magnification $\sigma_{\mu_{\rm res}}$ as a function of the siren's redshift $z$ under different observational schemes. Solid: the dispersion without correction ($\mu_{\rm est} = 1$). Dashed: the residual dispersion with convergence reconstructed from a JWST-like deep-field survey. Dotted: the residual dispersion with convergence reconstructed from a hybrid observation consisting of a JWST-like deep-field survey and an LSST-like wide-field survey. Dot-dashed: same as the dashed case but free from the mass-sheet degeneracy problem. The filter scale in all cases is tuned to be optimal.}
  \label{fig:ms_redshift_all}
\end{figure}

With the optimal filter scale $\theta_s$, Fig.~\ref{fig:ms_redshift_all} illustrates the dispersion of the residual magnification $\sigma_{\mu_{\rm res}}$ as a function of the siren's redshift $z_s$ under different observation schemes. The weak-lensing errors can be reduced by half for sirens at $z_s = 2.9$ on average under a futuristic hybrid observation (JWST-like + LSST-like surveys). A similar but smaller reduction is achievable for sirens in the range $1<z<3$. It is clear from this Figure that the mass-sheet degeneracy problem in a JWST-like deep-field survey is not dominant, since the total elimination of mass-sheet degeneracy by an (artificial) perfect wide-field survey does not improve the results significantly, and an LSST-like wide-field survey already produces results close to the perfect case and hence is good enough for breaking the mass-sheet degeneracy in a JWST-like deep-field survey. This conclusion may change if the total area of the adopted deep-field survey deviates from the JWST UDF configuration, and the galaxy density of the chosen wide-field survey is also important to this conclusion.

\begin{figure}
  \graphicspath{ {./ZF_fig/} }
  \includegraphics[width=\linewidth]{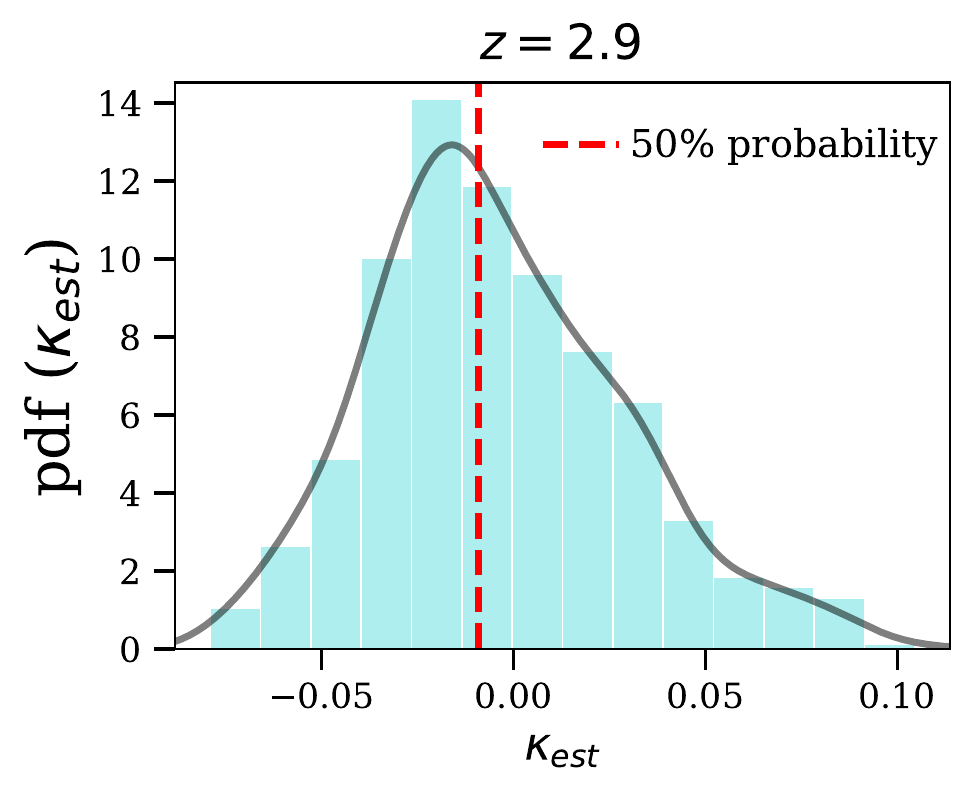}
  \caption{The probability distribution of estimated convergence $\kappa_{\rm est}$ at $z = 2.9$. The black solid curve shows the corresponding probability density function. The red dashed line denotes the median of the estimated convergence which is clearly less than zero.}
  \label{fig:prob_ka_new}
\end{figure}

\begin{figure}
  \graphicspath{ {./ZF_fig/} }
  \includegraphics[width=\linewidth]{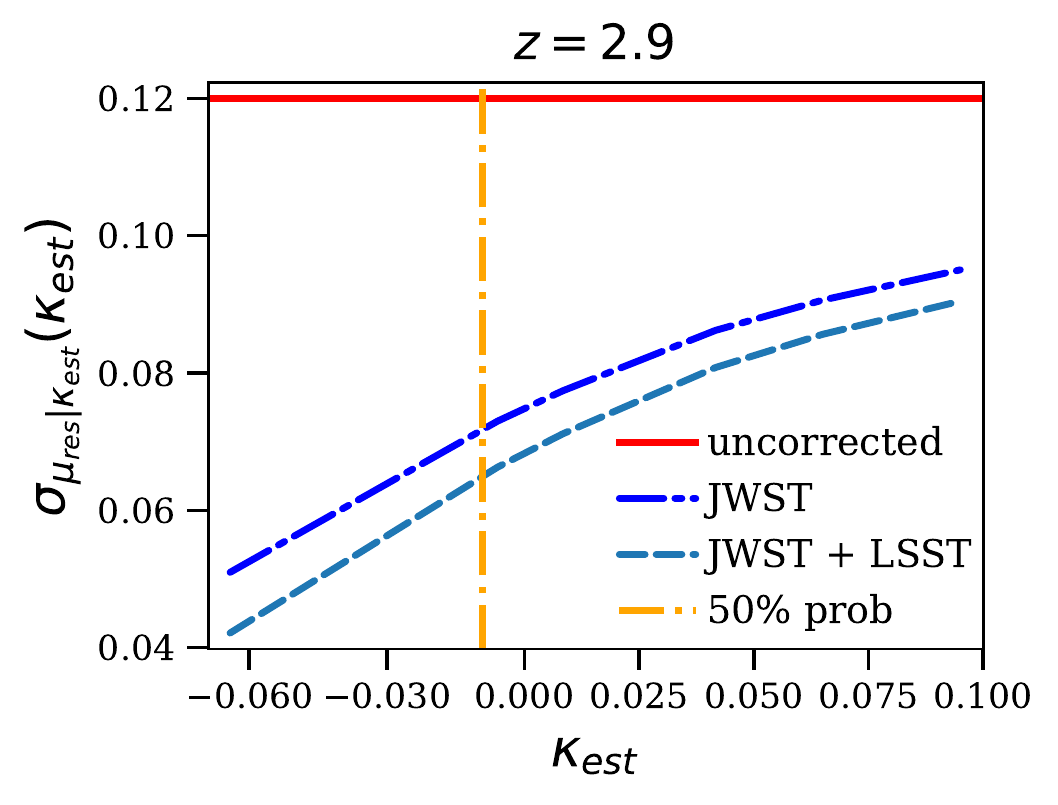}
  \caption{The dispersion $\sigma_{\mu_{\rm res}|\kappa_{\rm est}}(\kappa_{\rm est})$ of the conditional distribution $p(\mu_{\rm res}|\kappa_{\rm est})$ of the residual magnification $\mu_{\rm res}=\mu - \mu_{\rm est}$ as a function of the estimated convergence $\kappa_{\rm est}$ at redshift $z = 2.9$. Different observation schemes are considered and the vertical dashed line is the same as in Fig.~\ref{fig:prob_ka_new}. The solid red line indicates the dispersion of the uncorrected magnification. There is a 50\% probability that the weak-lensing error for a siren at $z_s = 2.9$ can at least be reduced by half and the reduction can reach $\sim 65\%$ in the most favourable case.}
  \label{fig:cond_ul}
\end{figure}

In a Bayesian approach, one might prefer to use the individual probability distributions $p(\mu_{\rm res}|\kappa_{\rm est})$ of the residuals $\mu_{\rm res}$ given the individual estimate of convergence $\kappa_{\rm est}$ rather than using a common residual distribution $p(\mu_{\rm res})$. The probability distribution of estimated convergence $p(\kappa_{\rm est})$ at $z = 2.9$ is represented in Fig.~\ref{fig:prob_ka_new}. It is clear that the median is less than zero, i.e. it is more probable to obtain an estimated convergence less than zero. Fig.~\ref{fig:cond_ul} quantifies the value of $\sigma_{\mu_{\rm res}|\kappa_{\rm est}}$ as a function of the estimated convergence $\kappa_{\rm est}$ from different observation schemes at $z = 2.9$. As Fig.~\ref{fig:cond_ul} depicts, low estimated convergence corresponds to small residual dispersion, which appears more encouraging than the marginalized residual dispersion. The vertical dashed line indicates the median of $\kappa_{\rm est}$, which is the same as the dashed line in Fig.~\ref{fig:prob_ka_new}. Therefore, under a futuristic hybrid observation scenario, there is a 50\% probability that the weak-lensing error for a siren at $z = 2.9$ can at least be reduced by half and the reduction can reach $\sim 65\%$ in the most fortuitous case. 

\section{Discussion}\label{Discussion}
In Sec.~\ref{hybrid}, the main conclusion presented is that, even for sirens at $z_s = 2.9$, the weak-lensing errors can be reduced by about a factor of two on average using convergence maps reconstructed from galaxy shape information, provided we have access to future hybrid observations that combine wide- and deep-field surveys. Similar results were obtained in \citet{shapiroDelensingGravitationalWave2010b} but with a much lower expectation on the depth of the deep-field survey. The deep-field survey adopted in \citet{shapiroDelensingGravitationalWave2010b} has galaxy density $n_{\rm gal} = 1000\,{\rm arcmin}^{-2}$ while the JWST-like UDF adopted in our work assumes a galaxy density $n_{\rm gal} \approx 2500\,{\rm arcmin}^{-2}$. Moreover, \citet{shapiroDelensingGravitationalWave2010b} only assumed 2D galaxy shape maps while we make use of the redshift information for individual galaxies. The reason why \citet{shapiroDelensingGravitationalWave2010b} could achieve essentially an equivalent result but with simpler requirements is that some assumptions made in that work turn out to be invalid. 

First, \citet{shapiroDelensingGravitationalWave2010b} did not make use of any numerical N-body simulations to obtain the matter fluctuation distribution, instead extending the \citet{smithStableClusteringHalo2003} fitting formula for the matter power spectrum beyond its accepted range of validity and assuming that matter fluctuations remain Gaussian even at the smallest scales. This approximation should be tested by numerical N-body simulations like SLICS, which compute the non-linear evolution of dark matter under gravity and thus resolve the structure formation deep in the non-linear regime. 

\begin{figure}
  \graphicspath{ {./ZF_fig/} }
  \includegraphics[width=\linewidth]{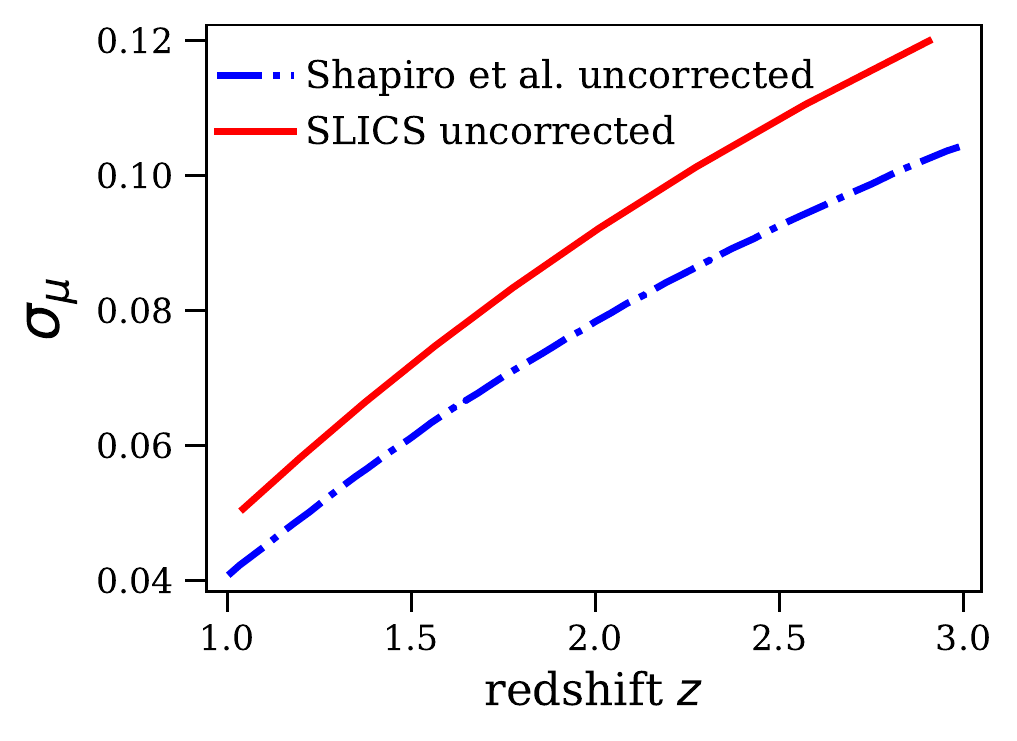}
  \caption{Comparison between the uncorrected magnification in \citet{shapiroDelensingGravitationalWave2010b} and in SLICS \citep{harnois-derapsCosmologicalSimulationsCombinedProbe2018}. The uncorrected magnification is larger in SLICS at all redshifts, indicating that \citet{shapiroDelensingGravitationalWave2010b} underestimated the small-scale fluctuations according to their assumptions.}
  \label{fig:shapiro}
\end{figure}

As Fig.~\ref{fig:shapiro} shows, the uncorrected magnification in \citet{shapiroDelensingGravitationalWave2010b} is smaller at all redshifts than the uncorrected magnification in SLICS. This indicates that the approximation and extension made in \citet{shapiroDelensingGravitationalWave2010b} appears to underestimate the matter power spectrum, especially at the small scales where the fitting formula is no longer valid. This also appears to render the shape noise problem less serious than it is in reality. Thus, the galaxy density required to achieve a factor of two reduction of the weak-lensing error appeared to be lower than it is found to be in our case.

\citet{shapiroDelensingGravitationalWave2010b} also underestimated the shear dispersion for a single galaxy $\sigma_\gamma$. They assumed $\sigma_\gamma = 0.2$, which incorporates intrinsic shape dispersion, background noise from the sky and the camera, and noise due to an imperfect deconvolution of the PSF. As the paradigm has shifted over the past decade, now the consensus is that the intrinsic shape dispersion for each galaxy is up to $\sigma_{\rm shape} = 0.26$, and higher for $\sigma_\gamma$ since other noises are included \citep{schaanLookingSameLens2017}.

Finally, \citet{shapiroDelensingGravitationalWave2010b} made use of the first weak gravitational flexion $\mathcal{F}$ and second weak gravitational flexion $\mathcal{G}$ maps to reconstruct the convergence maps. The weak gravitational flexions are derived from the third derivative of the lensing potential $\psi(\boldsymbol{\theta})$ and hence they are more sensitive to small-scale fluctuations than is shear. Including flexion allows one to supplement the shear measurements when reconstructing the small-scale modes. 

However, in practice, only the reduced flexion fields $F$ and $G$ are directly observable and measurements of the second reduced flexion $G$ are extremely delicate and generally dominated by noise \citep{roweFlexionMeasurementSimulations2013}. Furthermore, the reduced flexion dispersion per galaxy is underestimated in \citet{shapiroDelensingGravitationalWave2010b}. The first flexion dispersion per galaxy is assumed to be $\sigma_\mathcal{F} = 0.5/arcmin$ while a later paper demonstrated that $\sigma_{F} = 1.56/arcmin$ \footnote{The reduced factor is negligible in the weak-lensing regime.} inferred from the outcomes of \textit{Hubble’s Advanced Camera for Surveys} with realistic galaxy morphology \citep{roweFlexionMeasurementSimulations2013}. Other works based on the \textit{Hubble Space Telescope} (HST) including \citet{cainMEASURINGGRAVITATIONALLENSING2011} and \citet{lanusseHighResolutionWeak2016} reported or adopted the same value for $\sigma_{F}$.

In principle, $\sigma_{F}$ should be different in distinct deep-field surveys due to the unique PSF and PSF correction for each survey, but the values are expected to be around the same level. Further studies are needed to confirm this hypothesis. Nonetheless, given that the recent estimate of $\sigma_{F}$ for the \textit{Hubble} UDF is already three times larger than the estimate in \citet{shapiroDelensingGravitationalWave2010b}, we expect that the inclusion of flexion maps would not improve the result significantly and hence we have discarded the use of flexion in this analysis. Therefore, \citet{shapiroDelensingGravitationalWave2010b} overestimated the contribution from flexions $\mathcal{F}$ and $\mathcal{G}$ in weak-lensing reconstruction. All of the reasons above account for why \citet{shapiroDelensingGravitationalWave2010b} obtained similar delensing outcomes but with a much lower expectation value for the depth and number density of the deep-field survey. 

\citet{10.1111/j.1365-2966.2010.17963.x} also investigated the accuracy of weak-lensing reconstruction to infer the siren's magnification. They made use of galaxy shear and flexion maps to reconstruct the convergence maps, following the same assumptions for the single galaxy shape dispersion as in \citet{shapiroDelensingGravitationalWave2010b}. Therefore, \citet{10.1111/j.1365-2966.2010.17963.x} also underestimated the values for $\sigma_\gamma$, $\sigma_\mathcal{F}$ and $\sigma_\mathcal{G}$. Their analysis indicated that the weak-lensing error for a siren with $z_s = 3$ can be reduced to 72\% if only the shear data is available. 

Although \citet{10.1111/j.1365-2966.2010.17963.x} underestimated the single galaxy shape dispersion, they obtained worse error reduction results compared to our outcomes mainly because their adopted weak-lensing survey had a much lower galaxy number density, i.e. $n_{\rm gal} = 500\,{\rm arcmin}^{-2}$. 

Another possible reason for their worse performance is because of the adopted N-body simulation. The backbone dark matter N-body simulation used to generate the weak-lensing maps in \citet{10.1111/j.1365-2966.2010.17963.x} is the Millennium Simulation (MS) by \citet{2005Natur.435..629S}. While the comoving side length of a grid cube in the MS, $L_{\rm box}$, is comparable to SLICS, the MS has a smaller dark matter particle mass $m_p$ and a higher particle number $n_p$ within one comoving cube. However, the difference in $m_p$ and $n_p$ is not significant, so the resolution of MS is only slightly higher. The higher mass resolution in the backbone N-body simulation enables \citet{10.1111/j.1365-2966.2010.17963.x} to build weak-lensing maps with higher resolution. Therefore, the resolution of weak-lensing maps in \citet{10.1111/j.1365-2966.2010.17963.x} is $3.5$ arcsec, slightly higher than the resolution of $4.6$ arcsec of the weak-lensing maps in SLICS.

However, the adopted cosmological parameters in the MS are somewhat outdated: $\Omega_m = 0.25$, $\Omega_\Lambda = 0.75$, $h = 0.73$, $n_s = 1$ and $\sigma_8 = 0.9$, which differ from the recent best-fit values\footnote{See the end of Sec.~\ref{comb} for the best fit values from current observations.}. \citet{10.1111/j.1365-2966.2010.17963.x} used these parameters and simulated weak-lensing maps via the Multiple-Lens-Plane ray-tracing algorithm. Since the techniques and resolution are similar, the deviations in the weak-lensing maps are expected to be the consequence of differences in cosmology. These will lead to different initial amplitudes and evolution of the matter density fluctuations. Moreover, the comoving distance as a function of redshift will also be affected by cosmology, which plays a vital role in calculating the lensing properties at specific redshifts. Therefore, the cosmology dependence of the weak-lensing maps is rather complicated.

\begin{figure}
  \graphicspath{ {./ZF_fig/} }
  \includegraphics[width=\linewidth]{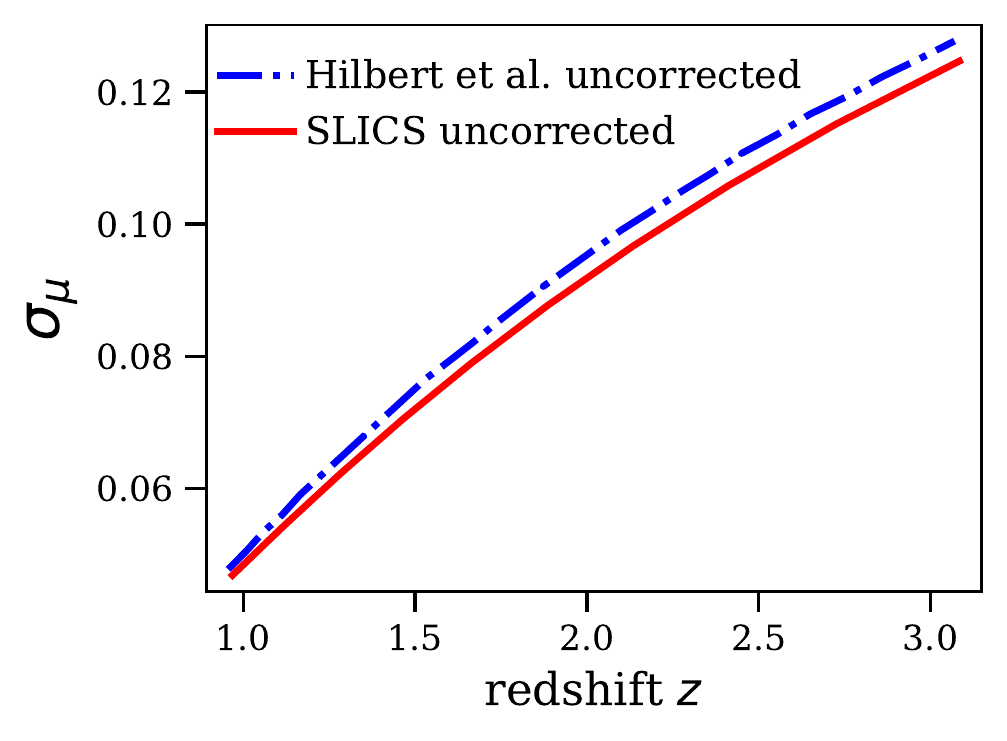}
  \caption{Comparison between the uncorrected magnification in \citet{10.1111/j.1365-2966.2010.17963.x} and in SLICS \citep{harnois-derapsCosmologicalSimulationsCombinedProbe2018}. The uncorrected magnification in \citet{10.1111/j.1365-2966.2010.17963.x} is slightly larger at all redshifts. The small deviation is expected to originate from differences in the cosmological model, as the adopted $\sigma_8$ is somewhat larger in \citet{10.1111/j.1365-2966.2010.17963.x}.}
  \label{fig:hil_com}
\end{figure}

\begin{figure}
  \graphicspath{ {./ZF_fig/} }
  \includegraphics[width=\linewidth]{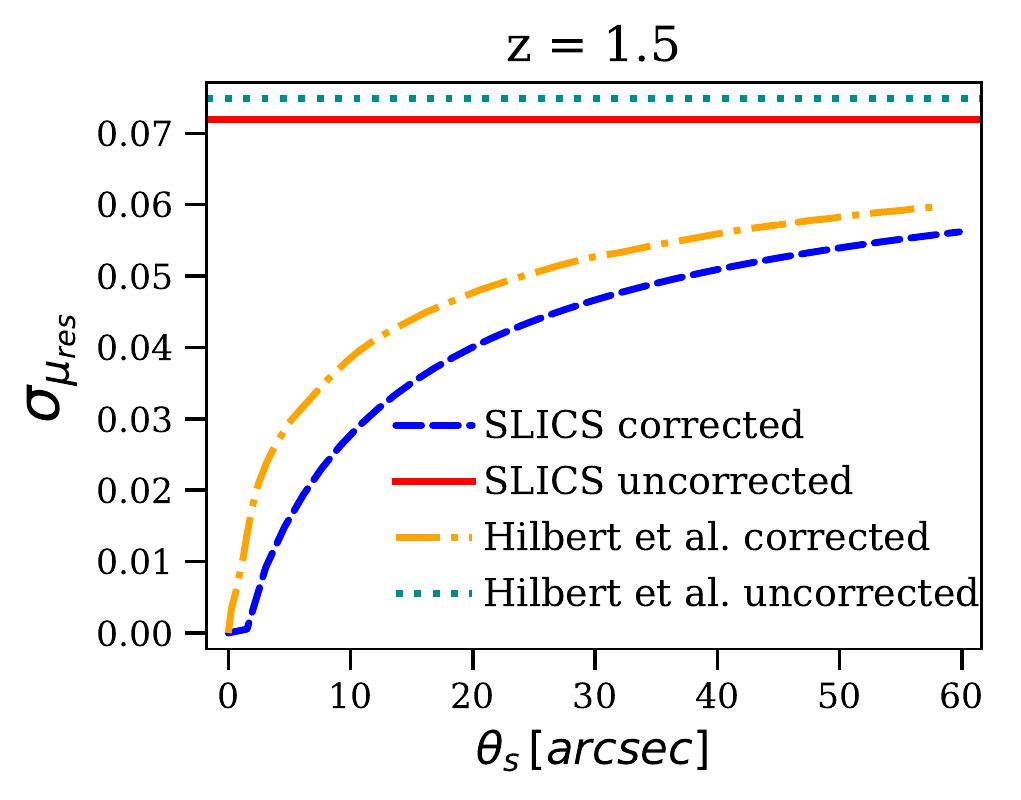}
  \caption{The dispersion of residual magnification $\sigma_{\mu_{\rm res}}$ at redshift $z = 1.5$ as a function of the smoothing filter scale $\theta_s$ from a perfect reconstruction in \citet{10.1111/j.1365-2966.2010.17963.x} (dot-dashed) and SLICS \citep{harnois-derapsCosmologicalSimulationsCombinedProbe2018} (dashed). Same as in Fig~\ref{fig:per_sm_2.9}, $\sigma_{\mu_{\rm res}}$ is contributed from the modes with scale below $\theta_s$. The horizontal lines mark the magnification dispersion without correction in \citet{10.1111/j.1365-2966.2010.17963.x} (dotted) and SLICS \citep{harnois-derapsCosmologicalSimulationsCombinedProbe2018} (solid). The dot-dashed line is higher than the dashed line, especially at small $\theta_s$, indicating that the convergence maps in \citet{10.1111/j.1365-2966.2010.17963.x} have larger small-scale fluctuations compared to SLICS.}
  \label{fig:per_sm_com}
\end{figure}

As shown in Fig~\ref{fig:hil_com}, the uncorrected magnification in \citet{10.1111/j.1365-2966.2010.17963.x} is slightly larger than the uncorrected magnification in SLICS at all redshifts, which is expected as the $\sigma_8$ is larger in \citet{10.1111/j.1365-2966.2010.17963.x}. Fig.~\ref{fig:per_sm_com} further demonstrates that the excess of the uncorrected magnification in \citet{10.1111/j.1365-2966.2010.17963.x} compared to SLICS is specifically contributed from small-scale fluctuations. Note that the shear measurements are less sensitive to small-scale fluctuations due to shape noise. Therefore, even if \citet{10.1111/j.1365-2966.2010.17963.x} could have adopted the same deep-field survey used in our work, we would anticipate that the outcomes would still be worse compared to ours. 

Surprisingly, we find that the statistical relations between lensing properties at different redshifts have similar trends regardless of the different cosmologies adopted in the two studies. This can be demonstrated by comparing Fig.~\ref{fig:correlation} and Fig.~\ref{fig:ratio_z} to the corresponding figures in \citet{10.1111/j.1365-2966.2010.17963.x} (Fig. 16 and Fig. 17). Therefore, the correlations between lensing quantities at different redshifts may be insensitive to cosmology, which would be interesting to investigate further in the future.

One more issue in \citet{10.1111/j.1365-2966.2010.17963.x} is the lack of simulations of wide-field surveys to break the mass-sheet degeneracy for the simulated deep-field survey. \citet{10.1111/j.1365-2966.2010.17963.x} assumed perfect removal of mass-sheet degeneracy by a wide-field survey, and hence the reduction of the weak-lensing errors should be even smaller in practice.

Finally, \citet{shapiroDelensingGravitationalWave2010b} and \citet{10.1111/j.1365-2966.2010.17963.x} studied only the homogeneous galaxy distribution at one specific redshift slice. By contrast, the galaxy catalogs used in our work consider both configurations of the observational program and the underlying matter distribution. Thus, the distribution is generally inhomogeneous at each redshift slice.
Therefore, we believe that the shape noise, which is related to the number of galaxies within one pixel, is better modelled in our work.

Furthermore, the selection effects in estimating lensing distributions for standard sirens was not investigated carefully in both \citet{shapiroDelensingGravitationalWave2010b} and \citet{10.1111/j.1365-2966.2010.17963.x}. The selection effects are induced by the correlation between the siren's total mass and the underlying mass density field since a high-mass siren will tend to be located in the denser part of the density field. Hence the lensing distributions for sirens with different total masses are distinct. However, \citet{shapiroDelensingGravitationalWave2010b} assumed that every line-of-sight in the observation field can serve as an equal sample towards the distributions of weak-lensing properties for the sirens. \citet{10.1111/j.1365-2966.2010.17963.x} made a similar assumption but weighted each line-of-sight by its inverse magnification to account for this selection effect. 

To model the selection effect more carefully, we made use of the empirical relation given in Eq.~\eqref{eq:empirical} to correlate the total mass of the SMBHB with the luminosity of the host galaxy in the \textit{r}-band. The luminosities of galaxies in the LSST-like catalogs are simulated by a HOD method and hence trace the underlying matter distribution and lensing fields. By selecting galaxies with a specific range of luminosities, we can model the selection effects and evaluate the dependence of lensing distributions on the siren's total mass. Nonetheless, a complete treatment using this approach will require future studies, which will be discussed further in the following section.
\vspace{-1em}
\section{Conclusions and Outlook}\label{Conclusion}
SMBHB systems are standard sirens that can provide accurate distance measurements based on their well-understood GW signals. However, gravitational lensing induces significant errors in the measured luminosity distances of high-redshift SMBHBs and the lensing errors are in general comparable to their measurement errors. Moreover, the lensing errors can not be averaged out due to the paucity of SMBHBs. Therefore, gravitational lensing of SMBHBs severely limits their usefulness as standard sirens and their power to constrain cosmological parameters. 

In this work, we investigated how much the weak-lensing errors can be reduced by weak-lensing reconstruction, making use of the latest numerical simulations. We then considered the potential impact of `delensing' strategies on cosmological parameter estimation. 

The main conclusion of our work is that the weak-lensing errors for sirens at $z_s = 2.9$ can be reduced by about a factor of two on average, under a futuristic hybrid observation involving wide- and deep-field galaxy surveys. However, such a hybrid observation requires an expensive ultra-deep field and hence might only be feasible for one particular siren in practice. Consequently, performing such a correction is unlikely to be worthwhile since the reduction of the error on the estimated luminosity distance for just one siren is likely insufficient to significantly improve the inference on the cosmological parameters.

Our analysis suggests that galaxy surveys can not sufficiently recover lensing structures at small scales due to the shape noise and thus the delensing performance is not satisfactory. Nonetheless, our conclusions might change if for example CMB lensing maps were incorporated. Other EM observations that can infer the density field along the trajectories of GW, such as tomographically-reconstructed galaxy density fields, might also be helpful for delensing. Future developments in the application of high-precision flexion measurements may also improve the outcomes significantly.

However, as pointed out at the end of Sec.~\ref{estimator}, we do not fully consider the non-Gaussian nature of the weak-lensing fields as we ignore the discussions about higher moments of the lensing distributions, which should be addressed in future work. Additionally, we have ignored higher-order weak-lensing effects such as reduced shear and flexion. The contaminating effects such as intrinsic alignments are also neglected in our work.

Another limitation of this work is related to the modelling of the selection effects. As demonstrated in Sec.~\ref{comb}, many high-redshift sirens do not have appropriate host galaxy candidates within the LSST-like catalogs since the suitable galaxies are below the luminosity threshold at the siren's redshift. This can be solved by constructing HOD-based deep-field catalogs from the same set of dark matter N-body simulations. Note that the mock deep-field catalogs in our work only simulate the galaxies in a simple way and do not contain information about the mass or luminosity of the galaxies. This is the reason why we do not consider the galaxies in the deep-field survey as SMBHB candidates and do not investigate the dependence of lensing distributions on the siren’s total mass. Therefore, a complete treatment of the selection effects should rely on future deep-field catalogs that simultaneously include weak-lensing maps and dim galaxies with the necessary properties.

Furthermore, the lensing convergence power spectra at different redshifts are related to cosmology. Therefore, the construction of an optimal magnification estimator should depend on the cosmology and the misalignment of the assumed cosmology to the real cosmology might also introduce a bias in the analysis. Additionally, the accuracy of the estimated magnification might also be cosmology-dependent. In this work, we only compare outcomes between two sets of cosmological parameters. The cosmology dependence of error reduction should be further explored in future studies. 

Besides, future studies might incorporate weak-lensing data directly into the estimation of cosmological parameters by making use of their cosmology dependence. The weak-lensing data can be obtained solely from the lensing of the GW signals \citep{congedoJointCosmologicalInference2019a} or derived from the synergy of galaxy surveys and GW experiments \citep{2022arXiv221006398B}. The cross-correlation of the GW lensing signals with the cosmic density field probed by EM observations has the potential to validate the standard model of cosmology and also the general theory of relativity \citep{2020MNRAS.494.1956M,PhysRevD.101.103509}.

Moreover, the effect of masks in the observed shear maps should also be accounted for properly. Since the KS inversion method is non-local, then realistic weak-lensing reconstructions suffer from the missing-data problem even if the targeted siren lying outside the mask. To solve the problem, methods equipped with inpainting techniques like the KS+ method can be helpful \citep{piresEuclidReconstructionWeaklensing2020}.

Finally, the weak-lensing reconstruction can be improved by new approaches for galaxy shear measurements such as the kinematic lensing method \citep{2022arXiv220911811S}. The kinematic lensing method combines photometric shape measurements with resolved spectroscopic observations to infer the intrinsic galaxy shape and directly estimate the lensing shear. The kinematic lensing method has an order of magnitude improvement over the traditional method, which is very useful in the context of weak-lensing reconstruction. Methods for 3D weak-lensing reconstructions might also help to reduce the weak-lensing errors for standard sirens \citep{leonardGLIMPSEAccurate3D2014}.
\vspace{-2em}
\section*{Acknowledgments}
\vspace{-0.5em}
The authors give our special thanks to Joachim Harnois-Deraps for providing detailed information on the SLICS catalogs and offering useful suggestions. We also thank Catherine Heymans, Ben Giblin, and David Bacon for useful discussion and insight. The authors are grateful for the siren catalogs provided by Liang-Gui Zhu and his team. M. H. is supported by the Science and Technology Facilities Council (Ref. ST/L000946/1). Finally, the authors acknowledge the many useful comments of the referee which have improved the clarity of this work.
\vspace{-2em}
\section*{Data Availability}
\vspace{-0.5em}
The data underlying this article will be shared on reasonable request to the corresponding authors.
\vspace{-2em}



\bibliographystyle{mnras}
\bibliography{reference_wzf} 




\onecolumn
\appendix

\section{Derivation of the posterior and likelihood}\label{appd}

The posterior on the cosmological parameters should be derived as,
\begin{equation*}
p(\Vec{\Omega}|\,\mathbf{D}_{\rm GW}\,, \mathbf{z}_{\rm s}\,,\mathbf{d}_{\rm lens}) = \frac{p(\Vec{\Omega}|\, \mathbf{z}_{\rm s}\,,\mathbf{d}_{\rm lens}) p(\mathbf{D}_{\rm GW}\,|\,\Vec{\Omega}, \mathbf{z}_{\rm s}\,,\mathbf{d}_{\rm lens})}{p(\,\mathbf{D}_{\rm GW}|\mathbf{z}_{\rm s}\,,\mathbf{d}_{\rm lens})}.
\end{equation*}
where the Bayesian evidence $p(\,\mathbf{D}_{\rm GW}|\mathbf{z}_{\rm s}\,,\mathbf{d}_{\rm lens})$ is ignored as it is irrelevant to the derivation. $\mathbf{z}_{\rm s}\,,\mathbf{d}_{\rm lens}$ can be regarded as background information.

The $\mathbf{D}_{\rm GW}$ includes the observed luminosity distance $D_L^{\rm obs}$ and their measurement uncertainties $\sigma_{D_L}$. In this paper, we only consider bright sirens, so we ignore the errors in redshifts since they are expected to be much lower than the errors in luminosity distance measurements. We also ignore the dependence of $\sigma_{D_L}$ on other parameters for simplicity. Then the likelihood becomes
\begin{align} \label{eq:prior}   
    p(\mathbf{D}_{\rm GW}|\,\Vec{\Omega}, \mathbf{z}_{\rm s},\mathbf{d}_{\rm lens}) 
    &\propto \prod_i p(D_L^{\rm obs}|\,\sigma_{D_L}, \Vec{\Omega}, z_{\rm s}, \mathbf{d}_{\rm lens})\\
   \intertext{We ignore the cosmology dependence of the weak-lensing data as assumed in Sec.~\ref{baye} and assume that weak gravitational lensing does not affect the measurement uncertainties $\sigma_{D_L}$\footnotemark. The observed luminosity distance is related to the lensing magnification by \ref{eq:1}. Then the likelihood is}
   &\propto \int \prod_i p(D_L^{\rm obs}|\,\sigma_{D_L}, \Vec{\Omega}, z_{\rm s}, \mathbf{d}_{\rm lens}, \mu)\,p(\mu|\,\mathbf{d}_{\rm lens}, z_{\rm s})\,d\mu \, ,\\
   \intertext{where $\mu$ is the true lensing magnification for each siren. The lensing data $\mathbf{d}_{\rm lens}$ can yield an estimate of the magnification $\mu_{\rm est}$ to correct the luminosity distance. Therefore,}
    &\propto \int \prod_i p(D_L^{\rm obs}|\,\sigma_{D_L}, \Vec{\Omega}, z_{\rm s}, \mu_{\rm est}, \mu)\,p(\mu|\,\mathbf{d}_{\rm lens}, z_{\rm s})\,d\mu \, .\\
   \intertext{Since the measurement uncertainties $\sigma_{D_L}$ are obtained from the Fisher matrix formalism with first-order approximation, the measurement error distribution should be Gaussian. The predicted luminosity distance $D_L$ (Eq.~\eqref{lum_z}) and observed luminosity distance $D_L^{\rm obs}$ for each siren should be corrected by the estimated magnification $\mu_{\rm est}$, then}
    &\propto\prod_i \int \exp\left[(D_L(\Vec{\Omega}, z_{\rm s}) / \sqrt{\mu}  \times \sqrt{\mu_{\rm est}} - D_L^{\rm obs}\times \sqrt{\mu_{\rm est}})^2/2\sigma_{D_L}^2 \right] \times p(\mu|\,\mathbf{d}_{\rm lens}, z_{\rm s})\, d\mu \,,
    \intertext{Since both $\mu$ and $\mu_{\rm est}$ are close to one with high probability (within 2$\sigma$), then $\sqrt{\mu_{\rm est}}/\sqrt{\mu} \approx (1 + (\mu_{\rm est} -\mu)/2)$. Let $D_L^{\rm cor}(\Vec{\Omega}, z_{\rm s}, \mu - \mu_{\rm est}) = D_L(\Vec{\Omega}, z_{\rm s}) \times (1 + (\mu_{\rm est} -\mu)/2)$. Finally, the likelihood is given by,}
    &\propto\prod_i \int \exp\left[(D_L^{\rm cor}(\Vec{\Omega}, z_{\rm s}, \mu - \mu_{\rm est}) - D_L^{\rm obs, cor})^2/2\sigma_{D_L}^2 \right] \times p(\mu - \mu_{\rm est}|\,\mathbf{d}_{\rm lens}, z_{\rm s})\, d\mu \,,
\end{align}\footnotetext{For high SNR signals, the ratio of the measurement errors $\sigma_{D_L}$ for the unlensed and lensed signals is equal to the ratio of their SNRs, by the Fisher Information Matrix formalism. The SNR of the original signal and the (de)magnified signal only differ by a factor of $\sqrt{\mu}$. Therefore, weak lensing will only induce a small change (probably within a few percent) in the measurement error $\sigma_{D_L}$ of the luminosity distance, which we regard as negligible.}
where $D_L^{\rm obs, cor} = D_L^{\rm obs} \times \sqrt{\mu_{\rm est}}$ and $p(\mu|\,\mathbf{d}_{\rm lens}, z_{\rm s})$ is equivalent to $p(\mu - \mu_{\rm est}|\,\mathbf{d}_{\rm lens}, z_{\rm s})$ since $\mu_{\rm est}$ depends only on $\,\mathbf{d}_{\rm lens}$ and $ z_{\rm s}$.


\bsp	
\label{lastpage}
\end{document}